\documentclass[acmsmall]{acmart}

\renewcommand\footnotetextcopyrightpermission[1]{} 
\setcopyright{none}

\usepackage{booktabs}   
\usepackage{subcaption} 

\usepackage{threeparttable}
\usepackage{makecell}
\usepackage{multirow}

\usepackage{listings}          
\usepackage[inference]{semantic}

\usepackage{dsfont}

\newcommand{\code}[1]{{\tt #1}}

\usepackage[normalem]{ulem}

\begin{document}

\title{Sparse Matrix Code Dependence Analysis Simplification at Compile Time}         


\author{Mahdi Soltan Mohammadi}
\affiliation{
  \department{Department of Computer Science}              
  \institution{University of Arizona}            
}
\email{kingmahdi@cs.arizona.edu}          

\author{Kazem Cheshmi}
\affiliation{
  \department{Department of Computer Science}              
  \institution{University of Toronto}            
}
\email{kazem@cs.toronto.edu}          

\author{Ganesh Gopalakrishnan}
\affiliation{
  \department{School of Computing}              
  \institution{University of Utah}            
}
\email{ganesh@cs.utah.edu}          

\author{Mary Hall}
\affiliation{
  \department{School of Computing}              
  \institution{University of Utah}            
}
\email{mhall@cs.utah.edu}          

\author{Maryam Mehri Dehnavi}
\affiliation{
  \department{Department of Computer Science}              
  \institution{University of Toronto}            
}
\email{mmehride@cs.toronto.edu}          

\author{Anand Venkat}
\affiliation{
  \institution{Intel Corporation}            
}
\email{anand.venkat@intel.com}          

\author{Tomofumi Yuki}
\affiliation{
  \department{}              
  \institution{Univ. Rennes, Inria, CNRS, IRISA}            
}
\email{tomofumi.yuki@inria.fr}          

\author{Michelle Mills Strout}
\affiliation{
  \department{Department of Computer Science}              
  \institution{University of Arizona}            
}
\email{mstrout@cs.arizona.edu}          


\begin{abstract}
Analyzing array-based computations to determine
data dependences is useful for many applications
including automatic parallelization, race detection, 
computation and communication overlap, verification, and shape analysis.
For sparse matrix codes, array data dependence
analysis is made more difficult by the use of
index arrays that make it possible to store only the nonzero entries of 
the matrix (e.g., in A[B[i]], B is an index array). 
Here, dependence analysis 
is often stymied by such indirect array accesses due to the
values of the index array not being available at compile time. 
Consequently, many dependences cannot be proven unsatisfiable 
or determined until runtime.
Nonetheless, index arrays in sparse matrix codes often have properties 
such as monotonicity of index array elements that can be exploited to
reduce the amount of runtime analysis needed.
In this paper, we contribute a formulation of array data dependence 
analysis that includes encoding  index array properties as 
universally quantified constraints.
This makes it possible to leverage
existing SMT solvers to determine
whether such dependences are unsatisfiable and
significantly reduces the number of dependences that
require runtime analysis in a set of eight sparse matrix kernels.
Another contribution is an algorithm for simplifying 
the remaining satisfiable data dependences by discovering
equalities and/or subset relationships.
These simplifications are essential to make
a runtime-inspection-based approach feasible.
\end{abstract}



\maketitle

\section{Introduction}

Data dependence analysis answers questions about which memory accesses 
in which loop iterations access the same memory location thus creating 
a partial ordering (or dependence) between loop iterations.
Determining this information enables 
iteration space slicing~\cite{Pugh97},
provides input to race detection, makes automatic parallelization 
and associated optimizations such as tiling or communication/computation 
overlap possible,
and enables more precise data-flow analysis, or abstract interpretation.
A data dependence exists
between two array accesses that (1) access the same array element with
at least one access being a write and (2) that access occurs within the loop
bounds for each of the accesses' statement(s).
These conditions for a data dependence has been posed 
as  a constraint-based decision problem~\cite{PW93}, a data-flow
analysis with polyhedral set information~\cite{Feau91}, and
linear memory access descriptors~\cite{Paek2002}.
However, such approaches require a runtime component when analyzing
codes with indirect memory accesses (e.g., A[B[i]], B being an index array)
such as those that occur in sparse matrix codes.
In this paper, we present an approach to improve the precision of
compile-time dependence analysis for sparse matrix codes
and simplification techniques for decreasing the complexity of
any remaining runtime checks.

\begin{figure}
\centering
\includegraphics[width=0.55\columnwidth]{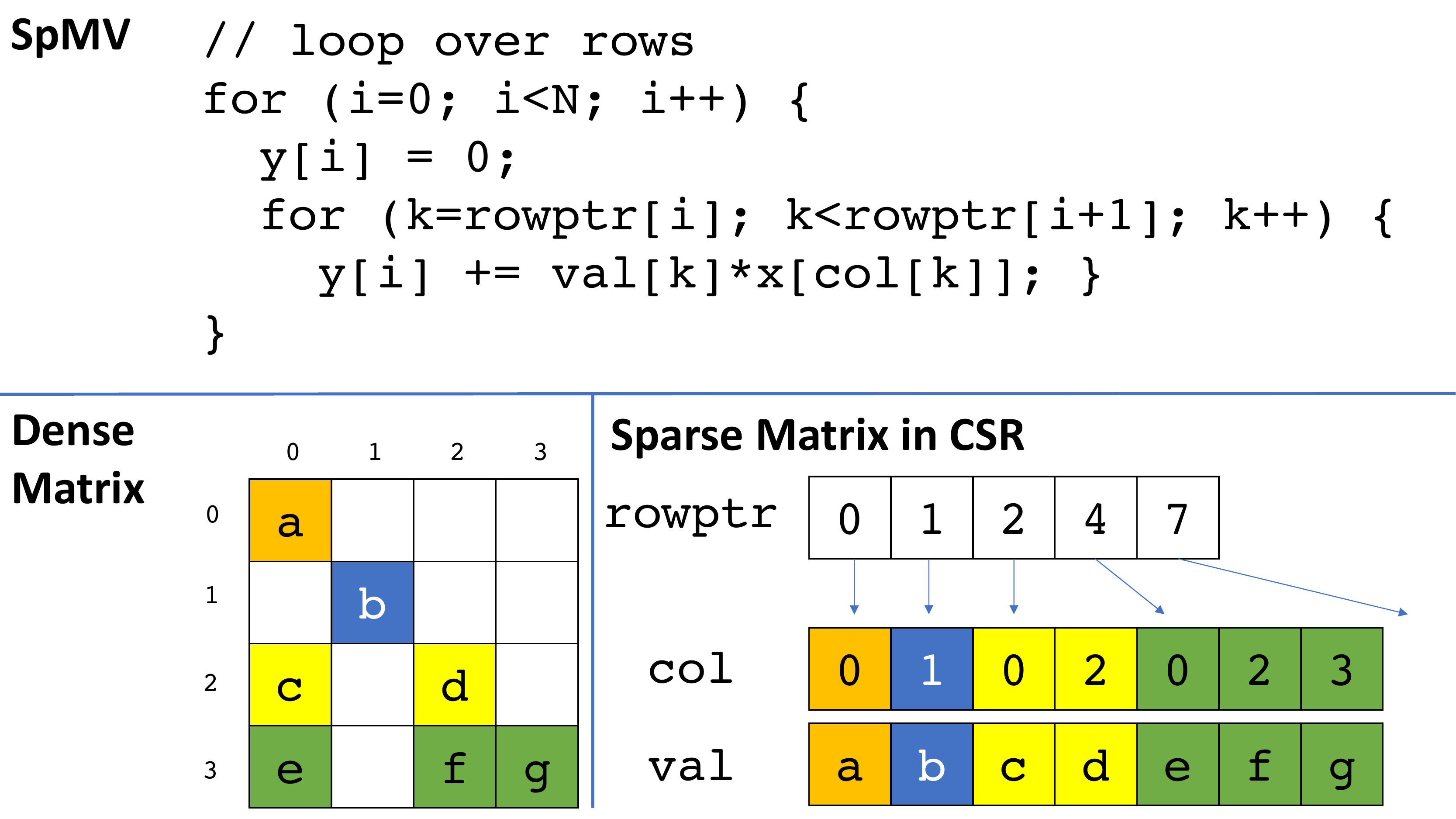}
\caption{Compressed Sparse Row (CSR) sparse matrix format. 
The {\tt val} array stores the nonzeros by packing each row in contiguous
locations.  The {\tt rowptr} array points to the start of each row in the {\tt
val} array. The {\tt col} array is parallel to {\tt val} and maps each nonzero
to the corresponding column.  }
\label{fig:csr}
\end{figure}

Sparse matrix computations occur in many codes, such as graph analysis,
partial differential equation solvers, and molecular dynamics
solvers. Sparse matrices save on storage and computation by only
storing the nonzero values in a matrix.
Figure~\ref{fig:csr}
illustrates one example of how the sparse matrix vector multiplication 
($\vec{y} = A\vec{x}$) can be written to use a common sparse matrix format 
called compressed sparse row (CSR). In CSR, the nonzeros are organized by row,
and for each nonzero, the column for that nonzero is explicitly stored.
Since the percentage of nonzeros in a matrix can be less than 1\%,
sparse matrix formats are critical for many computations to fit
into memory and execute in a reasonable amount of time.
Unfortunately, the advantage sparse matrices have on saving storage compared
to dense matrices comes with the cost of complicating program analysis. 
Compile-time only approaches resort to conservative 
approximation~\cite{barthou97fuzzy,Paek2002}.
Some approaches use runtime dependence tests to complement compile time
analysis and obtain more precise answers~\cite{PughWonn95,Rus2003,Oancea2012}. 
Runtime dependence information is also used to detect task-graph 
or wavefront parallelism that arises due to the 
sparsity of dependences~\cite{Saltz91,Rauchwerger95, Streit2015}.

The data dependence analysis approach presented here is constraint-based.
Some constraint-based data dependence analysis 
approaches~\cite{Nonlinear94,PughWonn95,pugh98constraintbased,Venkat:2016}
represent the index arrays in sparse matrix computations as uninterpreted functions 
in the data dependence relations. For example, the loop bounds for the {\tt k} loop in 
Figure~\ref{fig:csr} can be represented as $rowptr(i) \leq k < rowptr(i+1)$.
Previous work by others generates simplified constraints at compile time
that can then be checked at runtime 
with the goal of finding fully parallel loops~\cite{PughWonn95}.
We build on the previous work of~\cite{Venkat:2016}.
In that work, dependences in a couple of sparse matrix codes were
determined unsatisfiable manually, simplified using equalities found through
a partial ordering of uninterpreted function terms, or approximated by
removing enough constraints to ensure a reasonable runtime analysis complexity.
In this paper, we automate the determination of unsatisfactory data dependences, 
 find equalities using the Integer Set Library (ISL)~\cite{isl10}, 
have developed ways to detect data dependence subsets for simplifying runtime analysis, 
and perform an evaluation with eight popular sparse kernels.

\begin{figure}
\centering
\includegraphics[width=1\columnwidth]{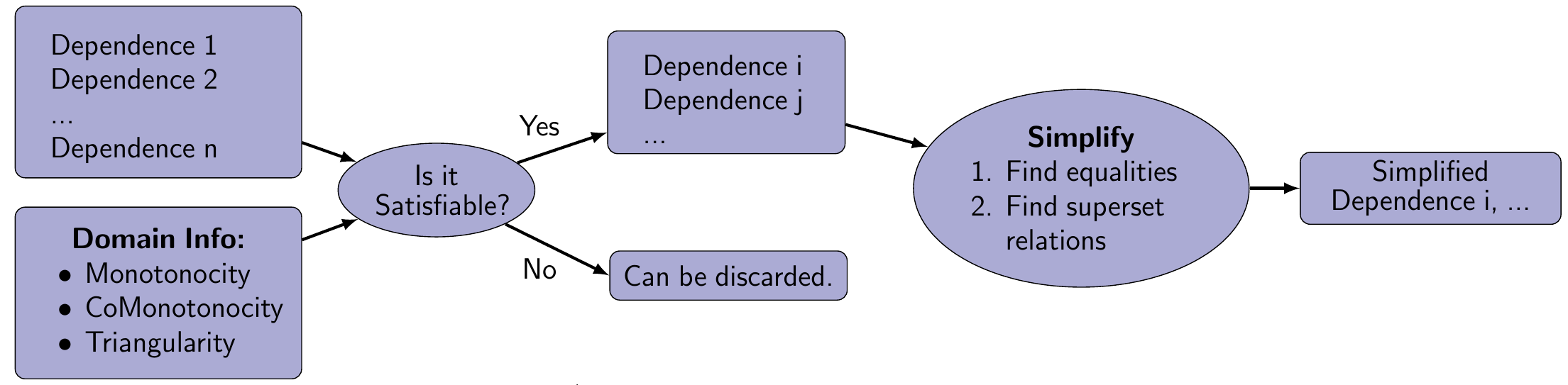}
\caption{This figure shows the overview of our approach
to eliminate or simplify dependences from sparse computation
utilizing domain information.
}
\label{fig:overview}
\end{figure}

\emph{In this paper, we have two main goals:
(1) prove as many data dependences as possible
to be \textit{unsatisfiable}, thus reducing the number of dependences 
that require runtime tests; and 
(2) simplify the satisfiable data dependences so that a
runtime inspector for them has complexity less than or equal
to the original code}.
Figure~\ref{fig:overview} shows an overview of our approach.
We use the ISL library~\cite{isl2018} like an SMT solver to determine which
data dependences are unsatisfiable. Next, we manipulate any
remaining dependence 
relations using IEGenLib~\cite{Strout16} and ISL libraries to discover equalities 
that lead to simplification.

Fortunately, much is known about the index arrays that 
represent sparse matrices as well as the assumptions made by 
the numerical algorithms that operate on those matrices. 
For example, in the CSR representation shown in Figure~\ref{fig:csr}, 
the {\tt rowptr} index array is monotonically strictly increasing.
In Section~\ref{unSatTheory} we explain how such information 
can be used to add more inequality and equality constraints 
to a dependence relation. The new added constraints in some cases
cause conflicts, and hence we can detect that
those relations are unsatisfiable.

The dependences that cannot be shown as unsatisfiable
at compile time, still require runtime tests.
For those dependences, the goal is to simplify the constraints and reduce 
the overhead of any runtime test. Sometimes index array properties 
can be useful for reducing the complexity of runtime inspector 
by finding equalities that are implied by the dependence constraints 
in combination with assertions about the index arrays.
The equalities such as $i = col(m')$ can help us remove
a loop level from the inspector, $i$ in the example. Since some equality
constraint would allow us deduce value of an iterator in the inspector from
another one, e.g, we can deduce $i$ values from $m'$ values using $i = col(m')$.
Another simplification involves determining when data dependence relations
for a code are involved in a subset relationship.  When this occurs,
runtime analysis need only occur for the superset.

This paper makes the following contributions:
\begin{enumerate}
\item An automated approach for the determination of unsatisfiable
 dependences in sparse codes.

\item An implementation of an instantiation-based decision procedure that discovers
equality relationships in satisfiable dependences.

\item An approach that discovers
subset relationships in satisfiable dependences thus reducing run-time
analysis complexity further.

\item A description of common properties of index arrays arising in sparse
matrix computations, expressed as universally quantified constraints.

\item Evaluation of the utility of these properties for determining
unsatisfiability or simplifying dependences from a 
suite of real-world sparse matrix codes.

\end{enumerate}

\section{Background: Data Dependence Analysis}
\label{sec:background}

Data dependence analysis of a loop nest is a common code analysis that is used
in different applications, such as automatic parallelization~\cite{Brandes88}
and data race detection~\cite{AtzeniGRALSLPM16}.
This section reviews the data dependence analysis process and how that process
differs when analyzing sparse matrix codes. Then, we review some of the applications
of data dependence analysis including an example of its use for
finding wavefront parallelism in sparse codes.

\subsection{Data Dependence Analysis}
\label{sec:datadep}

 A data dependence occurs between two
iterations of a loop when both of the iterations access the same memory
location and at least one of the accesses is a write.  Data dependence
constraints are of the following form:
\[
	Dep : (\exists \vec{I},\vec{I'})(\vec{I} \prec \vec{I'} \wedge F(\vec{I})=G(\vec{I'})
		\wedge Bounds(\vec{I}) \wedge Bounds(\vec{I'})),
\]
where $\vec{I}$ and $\vec{I'}$ are iteration vector instances from the same loop
nest, $F$ and $G$ are array index expressions to the same array
with at least one of the accesses being a write, and $Bounds(\vec{I})$ expands
to the loop nest bounds for the $\vec{I}$ iteration vectors.
In this paper, the term \emph{dependence relation} is
used interchangeably with dependence constraints by viewing them as a relation
between  $\vec{I}$ and $\vec{I'}$.

\begin{figure}
\begin{lstlisting}[columns=flexible,frame=single,keepspaces=true]
// forward solve assuming a lower triangular matrix.
for(i=0; i<N; i++) {
   tmp = f[i];
   for(j=0; j<i; j++) {
S1:   tmp -= A[i][j]*u[j];
   }
S2:u[i] = tmp / A[i][i];
}
\end{lstlisting}
\caption{Forward solve for a dense matrix.}
\label{lst:fsdense}
\end{figure}

For example, consider the dense matrix implementation for forward solve
in Figure~\ref{lst:fsdense}. Forward solve solves for the vector
$\vec{u}$ in the equation $A\vec{u}=\vec{f}$ assuming that the matrix is lower
triangular (i.e., nonzeros are only on the diagonals or below as shown in the
example in Figure~\ref{fig:csr}).  The dense forward solve code
has the following dependences for the outermost loop {\tt i}:
\begin{itemize}
\item A loop-carried dependence due to the scalar {\tt tmp}
variable.  However, since {\tt tmp} is written before being read in
each iteration of the {\tt i} loop, it is privatizable, 
which means each processor in a potential parallelization of the {\tt i} 
loop can be given its own private copy of {\tt tmp}.
\item A loop-carried dependence between the write {\tt u[i]}
in Statement {\tt S2} and the read {\tt u[j]} in Statement {\tt S1}
with constraints
\begin{align*} 
(\exists i,j, i',j')(&i<i' \wedge i=j' \wedge 0 \leq i,i' <N \wedge
0 \leq j <i \wedge 0 \leq j' <i').
\end{align*}

\end{itemize}

The iterators $i'$ and $j'$ are different instances of $i$ and $j$.
This dependence due to the writes and reads to array {\tt u} is satisfiable
because the computation for any row $i'$ depends on all previous rows $i< i'$.
This means that the outer loop in dense forward solve is fully ordered
due to data dependences
and therefore not parallelizable.

\subsection{Sparse Codes and Runtime Parallelism}
\label{sec:wavefront}

For sparse codes, after compile time dependence analysis, some remaining 
dependences may involve index arrays as subscript expressions.
The data dependence constraints can use uninterpreted functions to 
represent the index arrays at compile time.
Because the values of the index arrays are unknown until run time, 
proving such dependences are unsatisfiable may require runtime dependence testing.  
However, even when dependences arise at runtime, it still may be possible to implement
a sparse parallelization called wavefront parallelization.
Identifying wavefront parallelizable loops combines
compile time and runtime analyses.  The compiler generates inspector
code to find the data dependence graph at runtime.

We now consider the sparse forward solve with Compressed Sparse Row
CSR format in Figure~\ref{lst:FsCSR}. We are interested in detecting
loop-carried dependences of the outermost loop.  There are two pairs of
accesses on array {\tt u} in {\tt S1} and {\tt S2} that  can potentially cause
loop-carried dependences: {\tt u[col[k]]} (read), {\tt u[i]} (write); and {\tt
u[i]} (write), {\tt u[i]} (write). The constraints for the two dependence tests
are shown in the following.

\begin{figure}
\begin{lstlisting}[frame=single, columns=flexible]
// Forward solve assuming a lower triangular matrix.
for(i=0; i<N; i++) {
   tmp = f[i];
   for(k=rowptr[i]; k<rowptr[i+1]-1; k++) {
S1:   tmp -= val[k]*u[col[k]];
   }
S2:u[i] = tmp / val[rowptr[i+1]-1];
}
\end{lstlisting}
\caption{Forward solve for a sparse matrix in compressed
 sparse row (CSR).}
\label{lst:FsCSR}
\end{figure}

Dependences for the write/write {\tt u[i]} in {\tt S2}:
\begin{gather*}
(1) \;\; i = i' \wedge i < i' \wedge 0 \leq  i < N \wedge 0 \leq i' < N \; \\
\wedge rowptr(i) \leq k < rowptr(i+1) 
\wedge rowptr(i') \leq k'  < rowptr(k'+1)
\end{gather*}
\begin{gather*}
(2)\;\; i = i' \wedge i' < i \wedge 0 \leq  i < N \wedge 0 \leq i' < N \; \\
\wedge rowptr(i) \leq k < rowptr(i+1) 
\wedge rowptr(i') \leq k'  < rowptr(k'+1)
\end{gather*}

Dependences for read {\tt u[col[k]]} and write {\tt u[i]} in {\tt S1}, and {\tt S2}:
\begin{gather*}
(3)\;\; i = col(k') \wedge i < i' \wedge 0 \leq  i < N \wedge 0 \leq i' < N \;
\wedge \\ rowptr(i') \leq k' < rowptr(i'+1)
\end{gather*}
\begin{gather*}
(4)\;\; i = col(k') \wedge i' < i \wedge 0 \leq  i < N \wedge 0 \leq i' < N \;
\wedge \\ rowptr(i') \leq k' < rowptr(i'+1)
\end{gather*}

These dependences can be tested at runtime when concrete interpretations for
the index arrays (contents of arrays {\tt rowptr} and {\tt col}) are available.
The runtime dependence analyzers, called \textit{inspectors}~\cite{Saltz91},
may be generated from the dependence constraints~\cite{Venkat:2016}. 

\begin{figure}
\centering
\includegraphics[width=.5\columnwidth]{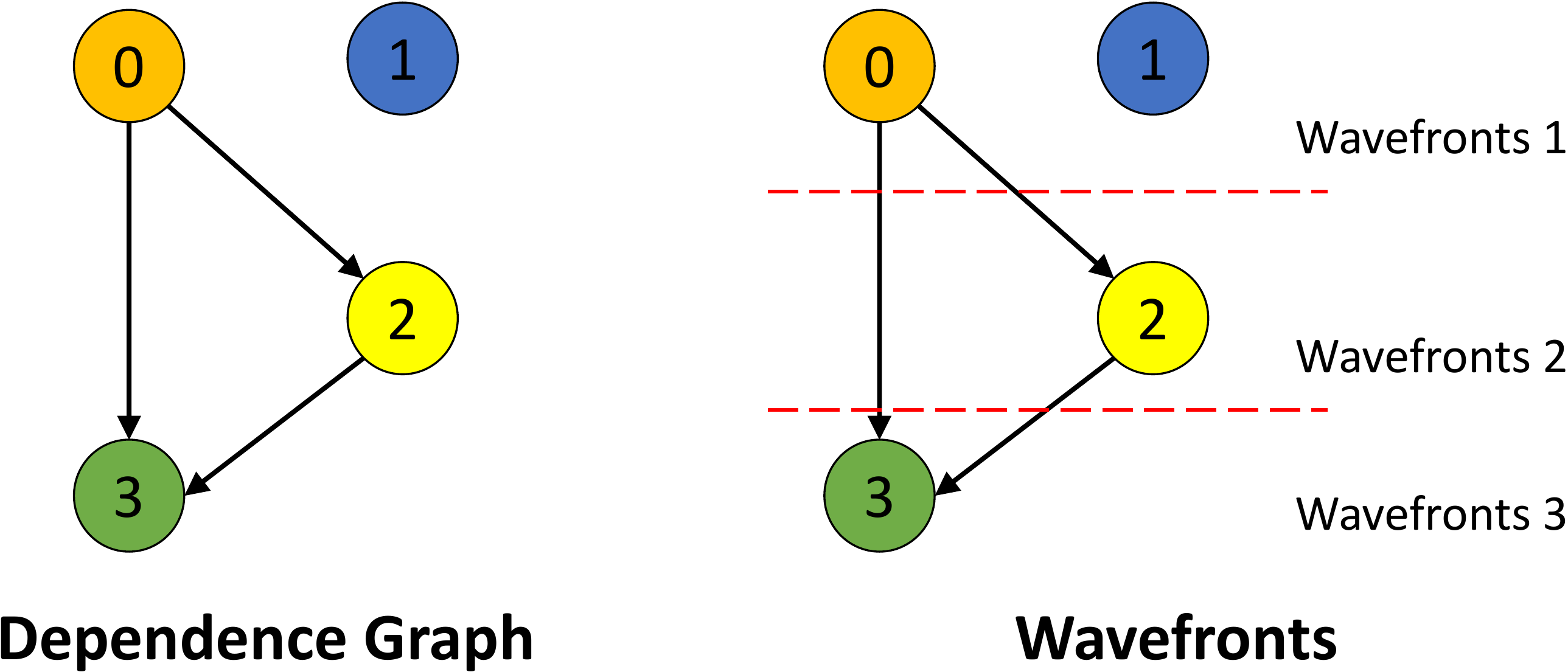}
\caption{Dependence graph for forward solve for sparse matrix
in Figure~\ref{fig:csr}.}
\label{fig:dep_graph}
\end{figure}

Suppose the matrix 
in the forward solve code in Figure~\ref{lst:FsCSR}
had the nonzero pattern as in Figure~\ref{fig:csr}. The runtime check 
would create the dependence graph for this example based on the
four dependences above as shown in Figure~\ref{fig:dep_graph}.
Once the dependence graph is constructed
a breadth-first traversal of the dependence
graph can derive sets of iterations that may be safely 
scheduled in parallel without a dependence violations, with each level set
being called a wavefront as shown in Figure~\ref{fig:dep_graph}.

\subsection{Applications of the Sparse Data Dependence Analysis}

Besides wavefront parallelism, there are many other uses for
sparse data dependence analysis.  Any application of sparse data dependence
analysis would benefit from a reduction in the number of data dependences
that need to be inspected at runtime and from any complexity reduction
of data dependences that do require runtime inspection.
Here we summarize some of those
applications.

{\bf Race detection:}
Dynamic race detection is an essential prerequisite to the
parallelization of existing sequential codes.
While the front-end static analysis methods employed
in these checkers~\cite{Atzeni2016} can often suppress race
checks on provably race-free loops, they fail
to do so when presented with non-affine access patterns
that occur in sparse matrix codes.
In addition to significantly increasing runtimes,
the shadow memory cells employed by dynamic race
checkers also increases memory pressure, often by a factor of four.
The techniques presented in this paper can help suppress
race checks when we can prove the independence of loop iterations.

{\bf Dynamic program slicing:} 
Pugh and Rosser introduced the concept of iteration space slicing where
program slicing is done on a loop iteration basis using 
Presburger representations~\cite{Pugh97}.
Similar dynamic approaches for tiling across loops in sparse codes 
were presented by various groups~\cite{dimeEtna00,StroutIJHPCA}.
All of these techniques would require runtime data dependence
analysis, thus disproving dependences or reducing the complexity
of inspecting dependences at runtime would be applicable.

{\bf High-level synthesis:}
Optimizations in high-level synthesis (HLS) uses runtime dependence checks.
In HLS, it is important to pipeline the innermost loops to get
efficient hardware.  Alle et al. have proposed using runtime dependence checks
to dynamically check if an iteration is in conflict with those currently in the
pipeline, and add delays only when necessary~\cite{Alle2013}.

{\bf Distributed memory parallelization:} 
Another possible application of our work can be found in the work by~\cite{Ravishankar2015}.
The authors produce distributed parallel code that uses MPI
for loops where there might be indirect loop bounds, 
and/or array accesses.
The read and write 
sets/data elements of each process are computed via an inspector 
where indirect accesses are involved to determine if each process is 
reading/writing data that is owned by other processes. 
Basumallik and Eigenmann use run-time inspection of data dependences
to determine how to reorder a loop to perform computation and communication
overlap~\cite{Basumallik06}.

\section{Automating (UN)Satisfiability Analysis for Sparse Data Dependences}

For any application of data dependence analysis for sparse codes,
the best outcome is to determine that a potential data dependence
is unsatisfiable.  Any dependence that is unsatisfiable does not
for runtime analysis.
Previous work used domain-specific knowledge about the index
arrays used to represent sparse matrices to guide manual determination
of unsatisfactory data dependences~\cite{Venkat:2016}.
In this paper, we show how to automate this process 
by specifying the domain-specific knowledge as universally
quantified constraints on uninterpreted functions and then
using instantiation methods similar to those used by 
SMT solvers to produce more constraints that can cause
conflicts.

\subsection{Detecting Unsatisfiable Dependences Using Domain Information}
\label{unSatTheory}

As an example of how domain information can be used to
show dependences are unsatisfiable consider the following 
constraints from a dependence relation:
$$i' < i\; \wedge\; k = m' \wedge\; 0 \le i,i' <n \wedge\;
rowptr(i) \leq k < rowptr(i+1) \wedge\; rowptr(i'-1) \leq m' < rowptr(i').$$
Some relevant domain information is that 
the {\tt rowptr} index array 
is strictly monotonically  increasing:
$$(\forall x_1,x_2)(x_1 < x_2 \implies rowptr(x_2) < rowptr(x_2)).$$

Since the dependence relation in question has the constraints
$i' < i$. Then, using the above strict monotonicity information 
would result in adding $rowptr(i') < rowptr(i)$.
But, considering  the constraint, $k = k'$, $rowptr(i) \leq k$, and
$m' < rowptr(i')$ we know that, $rowptr(i) < rowptr(i')$. 
This leads to a conflict, 
$$rowptr(i) < rowptr(i') \; \wedge \; rowptr(i') < rowptr(i).$$
This conflict would indicate the dependence in question was
unsatisfiable and therefore does not require any runtime analysis.

\subsection{Universally Quantified Assertions about Index Arrays}
Even if a formula that includes uninterpreted function calls is satisfiable in
its original form, additional constraints about the uninterpreted functions may
make it unsatisfiable. This has been exploited abundantly in
program verification community to obtain more precise results~\cite{Bradley2006,
Habermehl2008, Ge2009}. A common approach to express such additional
constraints is to formulate them as universally quantified assertions.  
For instance, \cite{Bradley2006} use following to indicate 
that array $a$ is sorted within a certain domain:

$$(\forall x_1, x_2)(i < x_1 \le x_2 < j \implies a(x_1) \le a(x_2) ).$$

There are several methods that SMT solvers use to reason about quantified
formulas, the most common one being quantifier
instantiation~\cite{Bradley2006,Ge2009,moura2007,
reynolds2014finding,reynolds2015counterexample,Loding2017}.
In quantifier instantiation, instances of universally quantified assertions, 
where the universally quantified variables are replaced with ground terms, 
are added to the original formula. 
Any of the added constraints might contradict constraint(s) in the formula
that would show the original formula is unsatisfiable.
For the general case of quantified first order logic, there is no complete 
instantiations procedure. That means the instantiation can go on forever
not exactly knowing whether the formula is satisfiable or unsatisfiable.
In some limited cases, the quantified assertions can be completely
replaced by a set of quantifier instances to construct an equisatisfiable
quantifier-free formula~\cite{Bradley2006, Ge2009}.

Combining the constraints from dependences with arbitrary universally
quantified assertions would create a first order 
logic theory that in general is undecidable. Undecidability would
imply that we cannot implement an algorithm for deciding the formulas
that would always terminate.
Numerous works such as~\cite{Bradley2006,Habermehl2008,Ge2009} present 
different decidable fragments of first order logic. 
The approach that these works use to make decidable fragments is to
put restrictions what type of universally quantified assertions
can be used. 
The restriction are usually on on syntax of the allowed 
assertions~\cite{Bradley2006,Habermehl2008},
and sometimes specific properties that a specific instantiation
procedure for assertions must have~\cite{Ge2009}.
We perform a terminating instantiation that is sound but incomplete.
In other words, the dependences we determine unsatisfiable
are in fact unsatisfiable, but we may characterize some unsatisfiable
constraints as may satisfiable.

\subsection{Domain Information about Index Arrays}
\label{sec:domain-assertions}
We represent domain information about index arrays as universally
quantified assertions. In this section, we illustrate 
some assertions relevant to numerical benchmarks
and relate the corresponding assertions to the existing
theory fragments.  
Table~\ref{tab:diffInfo} in Section~\ref{sec:eval} lists all the 
assertions we use in the evaluation.  Below are some example properties.

For the forward solve with compressed sparse
row (CSR) code in
Figure~\ref{lst:FsCSR}, we know the following:

\begin{itemize}
\item \textbf{Monotonic index arrays}: 
The row index array values increase monotonically.
This property of index arrays can be expressed with an
assertion about the uninterpreted function symbol
that represents the index array. 
For instance, in the example the $rowptr()$ function is monotonically increasing. 
If we assume that all the sparse matrix rows have at least one nonzero,
then $rowptr()$ is strictly monotonically increasing.
This assertion can be encoded as follows: 
	$$(\forall x_1,x_2)(x_1<x_2 \iff rowptr(x_1)<rowptr(x_2)).$$

\item \textbf{Lower Triangular Matrix}:
The forward solve algorithm shown in Figure~\ref{lst:FsCSR}
operates on lower triangular matrices.
For the CSR format that leads to the following domain-specific
assertion:
        $$(\forall x_1,x_2)(x_1<rowptr(x_2) \implies col(x_1)<x_2 )$$
This indicates that nonzeros of rows before row $i$ have columns less than $i$.
\end{itemize}

The domain information in Table~\ref{tab:diffInfo} in Section~\ref{sec:eval} 
can be represented with following 
general forms:

$$
1. \;\; (\forall x_1, x_2)(x_1+c_1 \le x_2 \implies f(x_1)+c_2 \le f(x_2)
$$
$$
2. \;\; (\forall x_1, x_2)(x_1+c_1 \le x_2 \implies f(x_1)+c_2 \le g(x_2)
$$
$$
3. \;\; (\forall x_1, x_2)(x_1+c_1 \le f(x_2) \implies g(x_1)+c_2 \le x_2) 
$$
Where $c_1$ and $c_2$ can be 0 or 1. 
The first and second assertions fit the decidable LIA fragment that is
presented by \cite{Habermehl2008}.  However, to the best of
our knowledge the third assertion form does not fit any previously presented
decidable fragment, and its decidability remains open.

Modern SMT solvers are equipped with heuristic-based quantifier instantiations
to reason about quantified formulas. Existing techniques for quantifier
instantiation can construct the set of instantiations for deciding some of our
assertions, e.g., non-strict monotonicity, but not for all of them.  For
unsatisfiable formulas with universal quantifiers where the solver only needs a
small set of relevant instances to find contradicting constraints, the existing
heuristics can work well. For all our examples, both Z3 and CVC4 were able to
identify all unsatisfiable dependences.  The solvers also time out for
satisfiable ones given a small timeout (5 seconds).  This is as expected, since
specific instances of universally quantified formulas usually do not help in
proving that the quantified formula is satisfiable.

Nonetheless, we cannot just use a conventional SMT solver like Z3 in
our context.
The key reason is that we are not just interested in satisfiability of the dependence
constraints.  If unsatisfiability cannot be proven statically, runtime checks
will be generated.  It is important for these runtime checks to be as fast as
possible, and hence we are also interested in using the assertions to decrease
the cost of runtime checks.  
For example, additional equalities means the data dependence inspector
iteration space
has lower dimensionality, thus reducing the algorithmic complexity of
runtime checks.  We illustrate the complexity reduction through instantiation
of assertions with two examples in Section~\ref{sec:simplifying-equality}.

\subsection{Detecting Unsatisfiable Sparse Dependences}
\label{sec:unsat-procedure}
Instantiation-based quantifier elimination is a natural choice for our context,
since we seek to either prove unsatisfiability or find additional constraints
that simplifies runtime checks. Unfortunately, our assertions are not fully
covered by decidable fragments~\cite{Bradley2006,Ge2009} where equisatisfiable
quantifier-free formulas can always be obtained.  Nonetheless, using
inspiration from the decidable fragments~\cite{Bradley2006,Ge2009} we have a
procedure that detects all unsatisfiable examples from our benchmark suite that
represent a wide range of numerical analysis codes. 

Note that we can show our general assertions (1), (2), and (3), presented
in Section~\ref{sec:domain-assertions} as:

$$\forall \vec{x},\; \varphi_{I}(\vec{x}) \implies \varphi_{V}(\vec{x})$$

Where $\vec{x}$ denotes vector of quantified variables,
$\varphi_{I}(\vec{x})$ denotes antecedent of the assertion,
and $\varphi_{V}(\vec{x})$ denotes consequent of the assertion.
Then the following definitions define our procedure to instantiate
quantified variables, 
and potentially use a SMT to detect their unsatisfiability.

\vspace{.1in}
\noindent
{\bf Definition 1 (E)} 
We define E to be the set of expressions used as arguments to 
all uninterpreted function calls in the original set of constraints.
We use this set to instantiate quantified assertions.

\vspace{.1in}
\noindent
{\bf Definition 2 ($UNSAT_\psi$)}
\begin{enumerate}
\itemsep2pt
\item The inference rules for turning the universally
quantified predicates into quantifier-free predicates is as follows:\\
\vspace{1pt}
\inference[\textsc{forall}] 
{
\psi [ \forall \vec{x},\; \varphi_{I}(\vec{x}) \implies \varphi_{V}(\vec{x}) ]
}
{
\psi [ \bigwedge_{\vec{x} \in E^n}    (\varphi_{I}(\vec{x}) \implies \varphi_{V}(\vec{x})) ]
}\\
\vspace{1pt}
where $E^n$ is the set of vectors of size $n=|\vec{x}|$ produced as Cartesian product of $E$.
\item Solve the quantifier-free formula $\psi$ output of step with an SMT solver that
decide union of quantifier-free theories of uninterpreted functions with equality 
and Presburger Arithmetics.
\end{enumerate}

\textbf{Correctness:}
Although the above procedure is incomplete, we do have soundness.
This means if a dependence is determined unsatisfiable, it in fact 
is not a dependence.  However, if a dependence is determined
satisfiable at compile time, it could be that at runtime the actual values of index
arrays lead to the dependence not being satisfiable.
Since our procedure is conservatively correct, it is sound.

To show that the decidability procedure $UNSAT_\psi$ is sound,
we need to show that if the original formula $\psi$ is 
satisfiable, then so is the unquantified formula $\psi'$,
	$$\psi \in SAT \implies \psi' \in SAT.$$
This is equivalent to
	$$ \psi' \notin SAT \implies \psi \notin SAT.$$
Since universal quantification is being replaced with 
specific expression instantiations to create $\psi'$,
$\psi'$ is a potentially weaker set of constraints
than $\psi$.
This means that $\psi'$ is a conservative approximation
of $\psi$.  As such, if $\psi'$ is not satisfiable,
then $\psi$ is not satisfiable.

\section{Simplifying the Dependences utilizing equalities}\label{sec:simplifying-equality}
The finite instantiation proposed in Section~\ref{sec:unsat-procedure} can
prove many of the dependence relations to be unsatisfiable. However, some of
the relations remain satisfiable, thus requiring runtime checks. It is then
important to minimize the runtime cost by simplifying the dependence relations
as much as possible. In this section, we discuss one of such simplifications
utilizing additional equalities after finite instantiations.

\subsection{Discovering New Equality Constraints and Their Usefulness}
Sometimes index array properties 
can help reduce the complexity of runtime inspectors
through introducing equalities to the dependence's constraints. 
The new equalities are discoverable after instantiating the
universally quantified assertions and combining those with other inequality
and equality relationships.
For instance, consider the following set of constraints; 
it is a satisfiable dependence that needs a runtime inspector with 
complexity of $O(n^2)$ to traverse the space of values for $i$ and $i'$:
$$(i \le i')  \wedge  (f(i') \le f(i)) \wedge (0 \le i,i' <n).$$
And assume we also know following universally quantified rule 
about the uninterpreted function $f$ (strict monotonicity):
$$(\forall x_1,x_2), (x_1 < x_2) \implies (f(x_1) < f(x_2)).$$
With any universally quantified implication, 
if the left side of the implication is true, then the right side must 
be true to satisfy the assertion (i.e., $p \implies q$). It is also 
the case that the contrapositive is true (i.e., $\neg q \implies \neg p$). 
For this example, the negation of the right-hand side of the implication is 
$f(x_2) \le f(x_1)$, which matches one of the constraints in the dependence.
Thus the negation of the left-hand side must be true and therefore 
$x_2 \le x_1$. With $x_1$ matching $i$ and $x_2$ matching $i'$, 
we find $i' \le i$. Thus an equality has been found:
$$ (i \le i' \wedge i' \le i) \implies  i = i' $$
Using this equality we can iterate over either $i$ or $i'$
in the inspector and calculate the other by taking advantage 
the equality. The runtime inspection would 
have complexity of only $O(n)$.

\begin{figure*}
\begin{lstlisting}[firstnumber=1,frame=tlrb,escapechar=|,
 mathescape=true, numbers=left, columns=flexible ]{Name}
for(int colNo = 0; colNo < n; ++colNo) {
   std::fill_n(f,n,0); //Zero initialization
   for(int nzNo = c[colNo]; nzNo < c[colNo + 1]; ++nzNo)
      f[r[nzNo]] = values[nzNo];
   for(int i = prunePtr[colNo], sw=0; i < prunePtr[colNo + 1]; ++i){
      for (int l = lC[pruneSet[i]], bool sw=false;; l < lC[pruneSet[i] + 1]; ++l){
          if (lR[l] == colNo && !sw) {
              tmp = lValues[l];
              sw=true;
          }
          if(sw){
S1:           f[lR[l]] -= lValues[l] * tmp;
          }
      }
   }
   if (f[colNo] <= 0) return false; //The matrix is not SPD
   lValues[lC[colNo]] = sqrt(f[colNo]);
   for(int j = lC[colNo] + 1; j < lC[colNo + 1]; ++j)
S2:   Values[j] = f[lR[j]] / sqrt(f[colNo]);
}
\end{lstlisting}
\caption{Static Left Cholesky code, which is a modified version of
Left Cholesky code~\cite{cheshmi2017sympiler}. }
\label{fig:lChol}
\end{figure*}

\subsection{Finding Equalities Example: Left Cholesky}
\label{sec:example}
For a more realistic example from one of the benchmarks used
in the evaluation, consider a maybe satisfiable dependence from 
our Static Left Cholesky shown in Figure~\ref{fig:lChol}. Following 
dependence is coming from a read in S1 (\code{lValues[l]}), and a
write in S2 (\code{lValues[j]}):

\begin{gather*}
\{ [colNo] \Rightarrow [colNo']\; : \; \exists j, i', l', (j = l') \wedge (colNo < colNo') \\
\wedge (0 \le colNo < n) \wedge (0 \le colNo' < n) 
\wedge (lcolptr(pruneSet(i')) \le l' < lcolptr(pruneSet(i')+1)) \\
\wedge (prunePtr(colNo') \le i' < prunePtr(colNo'+1)) 
\wedge (lcolptr(colNo) <  j < lcolptr(colNo+1)) \}
\end{gather*}

An inspector for this dependence is 
shown in Figure~\ref{leftChInsBefore}.
We do not need loops for $j$ and $l'$
in the inspector, because they can be projected out.
Note, index array \code{prunePtr} points to nonzeros in
the sparse matrix, ranging from 0 to number of nonzeros, $nnz$,
and $n$ denotes the number of column.
The two loops, \code{colNop} and \code{ip}, combined are traversing 
all the nonzero values and hence have a combined complexity of $nnz$, 
followed by the \code{colNo} loop traversing over columns, $n$. 
Consequently, the complexity of this inspector is $n(nnz)$.

The equality $colNo = pruneSet(i')$ is found using the additional knowledge
that {\tt lcolptr} is strictly monotonically increasing as demonstrated in the
following.

We have the following constraints in the original dependence:
\begin{align*}
	&lcolptr(pruneSet(i')) <= l' < lcolptr(pruneSet(i')+1)\\
        &\wedge \;\; j = lp \wedge lcolptr(colNo) < j < lcolptr(colNo+1),
\end{align*}
which gives the following through transitivity:
\begin{align*}
 & lcolptr(pruneSet(i')) < lcolptr(colNo+1)\;\; \\ 
 &\wedge\;\; lcolptr(colNo) < lcolptr(pruneSet(i')+1).
\end{align*}

We have the following assertion:
	$$(\forall x_1,x_2)(lcolptr(x_1) < lcolptr(x_2) \implies x_1 < x_2)$$
where two instances $x_1=pruneSet(i'),x_2=colNo+1$ and $x_1=colNo,x_2=pruneSet(i')+1$ give new constraints:
   \begin{align*}
&pruneSet(i')<colNo+1 \wedge colNo<pruneSet(i')+1\\
\Rightarrow \;\;  &pruneSet(i')\le colNo \wedge colNo\le pruneSet(i')\\
\Rightarrow \;\; &colNo = pruneSet(i')
\end{align*}

The optimized inspector based on new discoveries
is shown in Figure ~\ref{leftChInsAfter}.
We do not need loop for $colNo$, since we can get its values from 
$pruneSet(i')$ based on
$colNo = pruneSet(i')$.
This simplified inspector
has a complexity of $(nnz)$,
 which is significantly better than the original $n(nnz)$.

\begin{figure}
\noindent\begin{subfigure}{.48\textwidth}
\begin{lstlisting}[firstnumber=1,frame=tlrb,escapechar=|,
 mathescape=true, numbers=left ]{Name}
for(colNop = 0; colNop <n ; colNop++)
   for(ip = prunePtr(colNop); 
      ip < prunePtr(colNop+1); ip++) {
      for(colNo=0; colNo<n; colNo++) {
         if(lcolptr(colNo) < lcolptr(colNo+1) && ...) 
            // Add a flow dependence between colNo and ColNop
    }
  }
\end{lstlisting}
\caption{Inspector with the original dependence constraints.}
\label{leftChInsBefore}
\end{subfigure}\hfill
\begin{subfigure}{.48\textwidth}
\begin{lstlisting}[frame=tlrb,
                numbers=left,
                escapechar=|,
                mathescape=true, 
                firstnumber=1]{Name}
for(colNop = 0; colNop <n ; colNop++)
   for(ip = prunePtr(colNop); 
      ip < prunePtr(colNop+1); ip++) {
      colNo = pruneSet(ip);
      if(lcolptr(colNo) < lcolptr(colNo+1) && ...) 
         // Add a flow dependence between colNo and ColNop

  }
\end{lstlisting}
\caption{Inspector with an additional equality: $colNo = pruneSet(i')$.}
\label{leftChInsAfter}
\end{subfigure}
\caption{Inspector pseudo-code for dependence constraints 
in Section~\ref{sec:example}, before and after utilizing index array properties
to add new equalities.  We obtain the equality $colNo = pruneSet(i')$
using the properties as described in Section~\ref{sec:example}. Notice how this
equality is used to remove loop iterating over {\tt i} in Line 3.}
\label{fig:lCholins}
\end{figure}

\section{Simplifying the Dependences Utilizing Superset Relationship}
\label{sec:simplifying-superset}

Another way to deal with data dependence relations that cause complex
runtime analysis is to remove it from consideration by determining it is a
 subset of a less expensive relation.
Consider two dependence relations $R1$ and $R2$, and two iterations of the
outermost loop $i$ and $i'$. If we can show that for all $i$ and $i'$ that are
dependent according to $R2$, the same pairs of $i$ and $i'$ are also dependent
according to $R1$, then it is sufficient to only test $R1$. We say that $R1$ is
a superset of $R2$, written $R1 \supseteq R2$, in such cases, and remove $R2$
from runtime check. Note that in the above definition, $R1$ may have more pairs
of outermost iterators that are dependent than $R2$.

Taking advantage of this redundancy can result in lower complexity runtime analysis. 
As an example, consider the  Incomple Cholesky code shown in
Figure~\ref{fig:IC0_CSC}. In section, we refer to an array access $A$ at statement $S$ as
$A@S$ for brevity. One of the dependence tests is between the write
\code{val[k]@S3} and the read \code{val[m]@S3}. This test is
redundant with the test between the write \code{val[k]@S3} and the read
\code{val[m]@S2}. This is because an iteration of the \code{i} loop
that make the read from \code{val[m]} in \code{S3} is guaranteed to access the same
memory location while executing the loop surrounding \code{S2}. Thus, the more
expensive check between accesses in \code{S3} can be removed.

\begin{figure}
\begin{lstlisting}[columns=flexible,firstnumber=1,frame=tlrb,escapechar=|,
 mathescape=true, numbers=left ]{Name}
for (i = 0; i < n; i++) {
S1: val[colPtr[i]] = sqrt(val[colPtr[i]]);

   for (m = colPtr[i] + 1; m < colPtr[i+1]; m++)
S2:    val[m] = val[m] / val[colPtr[i]];

   for (m = colPtr[i] + 1; m < colPtr[i+1]; m++) 
      for (k = colPtr[rowIdx[m]] ; k < colPtr[rowIdx[m]+1]; k++)
         for ( l = m; l < colPtr[i+1] ; l++)
            if (rowIdx[l] == rowIdx[k] && rowIdx[l+1] <= rowIdx[k])
S3:            val[k] -= val[m]* val[l];
}
\end{lstlisting}
\vspace{-0.3cm}
\caption{\label{fig:IC0_CSC} Incomplete Cholesky0 code from 
SparseLib++~\cite{pozo1996sparselib++}. Some variable names have been changed.
The arrays \code{col} and \code{row} are to represent common colPtr, and rowIdx in CSC format.}
\vspace{-0.3cm}
\end{figure}

In this section, we describe our approach to identify redundant dependence
relations. The key challenge is to determine superset relations between two
dependence tests involving uninterpreted functions. We present two approaches
that cover important cases, and discuss possible extensions.

\subsection{Trivial Superset Relations}
Given a polyhedral dependence relation, it is easy to characterize the pairs of
loop iterations that are dependent. All the indices that do not correspond to
the loop iterators in question can be projected out to obtain the set of
dependent iterations. These sets can be compared to determine if a dependence
test is subsumed by another.  In principle, this is what we do to check if a
dependence relation is redundant with another. However,  dependence relations
from sparse codes
may have variables passed as parameters to uninterpreted functions.
Such variables  cannot be projected out.
Thus, we employ
an approach based on similarities in the constraint systems. The trivial case
is when the dependence relation $R1$ is expressed with a subset of constraints
in another relation $R2$. If this is the case, then $R1$ can be said to be
superset equal to $R2$.

We illustrate this approach with the earlier example from Incomplete Cholesky.
We take two dependence relations, $R1$ between \code{val[k]@S3}
and \code{val[m]@S2}, and $R2$ between \code{val[k]@S3} and
\code{val[m]@S3}. The relations---omitting the obviously common constraints 
for \code{val[k]@S3}---are:
\begin{gather*}
R1 = \{ [i,m,k,l] \rightarrow [i',m'] : k = m' \wedge i < i' \wedge 0 \le i' < n \wedge \; col(i')+1 \le m' < col(i'+1) \}\\
R2 = \{ [i,m,k,l] \rightarrow [i',m',k',l'] : k = m' \wedge i < i' \wedge 0 \le i' < n \wedge \; col(i')+1 \le m' < col(i'+1) \\
\wedge \; col(row(m'))\le k' < col(row(m)+1) \wedge m'\le l' < col(i+1)\\
\wedge \; row(l')=row(k') \wedge row(l'+1)\le row(k') \}
\end{gather*}

Since $R1$ is expressed with a subset of constraints in $R2$, 
we may conclude that $R1 \supseteq R2$.

\subsection{Superset Relation due to Overlapped Accesses}
\label{sec:OverlappedSuperset}

The trivial check is sufficient for many pairs of relations. However, some
relations require a more involved process. We use a different dependence
relation from Incomplete Cholesky (Figure~\ref{fig:IC0_CSC}) as an example of
such cases.  We consider the dependence relation $R3$ between \code{val[k]@S3}
and \code{val[l]@S3} that is redundant with $R1$. This can be intuitively
observed from the code structure: the set of memory locations that may be
accessed by the read of \code{val[l]} when $l=m$, i.e., the first iteration of
the $l$ loop, is exactly the same as the reads by \code{val[m]@S2}.  This
guarantees that even if the guard on \code{S3} always evaluated to true, the
dependence between iterations of the \code{i} loop would be redundant with that
imposed by \code{S2}.

The constraints for $R3$ (omitting those for \code{val[k]@S3}) are as follows:
\begin{gather*}
R3 = \{ [i,m,k,l] \rightarrow [i',m',k',l'] : k = l' \wedge i < i' \wedge 0 \le i' < n \wedge \; col(i')+1 \le m' < col(i'+1)\\
\wedge \; col(row(m'))\le k' < col(row(m)+1) \wedge m'\le l' < col(i+1)\\
\wedge \; row(l')=row(k') \wedge row(l'+1)\le row(k') \}
\end{gather*}

\begin{enumerate}
\item We first identify that $k=m'$ in $R1$ is not a constraint in $R3$.
\item The equality $k=m'$ has a similar (one side of the equality is the same) equation $k=l'$ in $R3$.
\item The bounds on $m'$ and $l'$ are collected from the respective constraints.
\item Because the bound on $m'$ subsumes that of $l'$, and since $k=m'$ was the
only constraint that was not in $R3$, we may conclude that  $R1 \supseteq R3$.
\end{enumerate}

It is important to note that the bounds on $l'$---the set of values accessed in
the subset relation---can be conservative, i.e., \emph{may} accesses, but the
bounds on $m'$---the set of values accessed in the superset relation---must be
exact. If both bounds represent may accesses, then the superset relation does
not hold. This is important for situations as illustrated in the example above,
where statements have data-dependent guards.

\subsection{Limitations and Extensions}
Although the approach presented above was able to cover all the important cases
we encountered, it is by no means complete. The difficulty of manipulating
integer sets with uninterpreted function symbols have led us to work directly
on the constraints. This may cause our approach to miss some superset
relations, since the same relation can be expressed in many different ways. Adding some
form of normalization to the constraint system will help us avoid such pitfalls.

The overlapped iterator approach to finding a superset in 
Section~\ref{sec:OverlappedSuperset} was developed specifically for the problematic
data dependence relation R3.  Future work includes developing a more general
simplification approach based on this overlapping iterator concept.

In terms of scaling, there  is potentially a 
problem of selecting the pairs of dependence relations to
test for redundancy. We currently try all possible candidate pairs, which does
not pose a problem since a large number of dependence relations are filtered
out through unsatisfiability test described in
Section~\ref{sec:unsat-procedure}. However, selecting promising pairs to limit
the number of tests would be an useful extension.

\section{Implementation}
The data dependence analysis and simplification have been automated 
except for the superset simplification.  
This section summarizes the software packages the implementation relies on, 
discusses some important optimization to make our implementation scalable,
and compares the ISL-based implementation with that of an SMT solver.

\subsection{Software Description}
The artifact for reproducing the results presented in this article
 can be found at the following public github repository:
https://github.com/CompOpt4Apps/Artifact-datadepsimplify-arXiv-July2018

We use three software packages to automate applying methods described 
in this paper: 
IEGenLib library~\cite{Strout16},
ISL library~\cite{isl2018},
and CHILL compiler framework~\cite{chill:ctop}.  
CHiLL is a source-to-source compiler framework for composing and applying
high level loop transformations to improve the performance of nested loop 
written in C. We utilize CHILL to extracted the dependence relations
from the benchmarks.
ISL is a library for manipulating integer sets and relations that 
only contain affine constraints.
It can act as a constraint solver by testing the
emptiness of integer sets. It is also equipped with a function for detecting
equalities in sets and relations.  ISL does not support uninterpreted
functions, and thus cannot directly represent the dependence constraints in
sparse matrix code. 
IEGenLib is a set manipulation library that can  manipulate 
integer sets/relations that  contain uninterpreted function symbols. 
The IEGenLib library utilizes ISL for some of its fundamental functionalities. 
We have implemented detecting unsatisfiable dependences and finding 
the equalities utilizing the IEGenLib and ISL libraries.

The following briefly describes how the automation works.
First, we extract the dependence relations utilizing CHILL, and
store them in IEGenLib. The user defined index array properties 
are also stored in IEGenLib.
Next, the instantiation procedure is carried out in IEGenLib.
Then inside IEGenLib, the uninterpreted functions are removed by replacing each
call with a fresh variable, and adding constraints that encode functional
consistency~\cite[Chapter 4]{Kroening2016}. 
Next, ISL can be utilized by IEGenLib to
find the empty sets, i.e, unsatisfiable relations.
Additionally, equality detection is available
as one of many operations supported by ISL
The finite instantiations described in
Section~\ref{sec:unsat-procedure} 
are intersections of the assertions with the dependence
relation. 

\subsection{Optimization}
A straightforward approach to implementing the procedure in
Section~\ref{sec:unsat-procedure} would be to take the  quantifier-free
formula resulting from instantiation,
replace the uninterpreted functions, and directly pass it to ISL.
However, this approach does not scale to large numbers of instantiations.
An instantiated assertion is encoded as a union of two constraints (${\neg}p
\vee q$).  Given $n$ instantiations, this approach introduces $2^n$
disjunctions to the original relation, although many of the clauses may be
empty.  In some of our dependence relations, the value of $n$ may exceed
$1000$, resulting in a prohibitively high number of disjunctions.  We have
observed that having more than 100 instantiations causes ISL to start
having scalability problems.

We apply an optimization to avoid introducing disjunctions when possible.
Given a set of instantiations, the optimization adds the instantiations to 
the dependence relation in two phases. 
The first phase only instantiates those that do not introduce disjunctions 
to the dependence relation. 
During this phase, we check if the antecedent is already part of 
the dependence constraint, and thus is always true. 
If this is the case, then $q$ can be directly added to the dependence relation. 
We also perform the same for ${\neg}q \implies {\neg}p$ 
and add ${\neg}p$ to the dependence relation if ${\neg}q$ is always true. 
The second phase adds the remaining instantiations that introduce disjunctions.
This optimization helps reducing the cost of dependence testing in two ways:
(1) if the relation is unsatisfiable after the first phase, disjunctions are completely avoided; 
and 
(2) the second phase only instantiates the remainder, reducing the number of disjunctions.

If the dependence relation remains non-empty after the second phase, then the
relation is checked at runtime. All equalities in a relation is made explicit
before inspector code generation with ISL so that the code generator can take
advantage of the equalities to simplify the generated code.

\subsection{Contrasting SMT with ISL}

SMT solvers are specialized for solving satisfiability problems expressed as a
combination of background theories. ISL is a library for manipulating integer
sets, and is specialized for the theory of Presburger arithmetic over integers.

The finite instantiation in Section~\ref{sec:unsat-procedure} is well-suited
for SMT solvers. In fact, SMT solvers are equipped with their own instantiation
algorithms that also work well for our dependence relations. However, SMT
solvers do not provide any equality relationships they might derive while answering
the satisfiability question.
Although it is possible to use SMT solvers to test if an equation is true for a
set of constraints, we cannot search for an equation given the constraints
(unless all candidates are enumerated---but there are infinite candidates in general). 

For our implementation, the choice was between adding finite instantiation to
ISL or adding equality detection to SMT solvers. We have chosen the former
option as it seemed simpler to do, and also because we are more familiar with
ISL.

\section{Evaluation of Unsatisfiability and Simplification Approaches}
\label{sec:eval}
In this section, we study the impact of our approach of utilizing domain information
about index arrays on the data dependence analysis of eight sparse kernels.
Our approach may help data dependence analysis in three ways: 
(1) The runtime check can be completely removed
if the dependences are proven unsatisfiable; 
(2) Deriving equalities from instantiated universally quantified
assertions about domain information 
can simplify dependences and reduce respected runtime check complexity; and
(3) Reducing all maybe satisfiable relations of a given code to
a set of dependence relations that encompass all potential dependences.
We do this by finding relations that are superset equal of other relations.
This can discard even more dependence relations that potentially
might need expensive runtime checks. 

We first describe the suite of numerical kernels that we have compiled to
evaluate our approach. Then we evaluate the impact of each step in our
approach, from the relevance of the index property assertions to the simplification
using superset relations. Finally, we report the complexity of inspectors
with and without our proposed simplifications.

\subsection{Numerical Algorithms in Benchmark}
\label{sec:benchmark}
We have included some of the most popular sparse kernels 
in a benchmark suite: (1) The Cholesky factorization, Incomplete LU0 and
Incomplete Cholesky0, and the sparse triangular solver, which are commonly used
in direct solvers and as preconditioners in iterative solvers; (2)  sparse
matrix vector multiplication, and
Gauss-Seidel methods, which are often used in iterative solvers.
Table~\ref{tab:suite} summarizes the benchmarks indicating which library each
algorithm came from and how the benchmark compares with the implementations in
existing libraries. 

\begin{table}
\begin{center}
\begin{threeparttable}
\caption{The benchmark suite used in this paper. The suite includes the
fundamental blocks in several applications. The suite is also selected to cover
both static index arrays, such as Gauss-Seidel, and dynamic index arrays, such
as Left Cholesky. The modification column shows the type of modification
applied to the original code.}
\begin{tabular}{| l | c | c | p{.35in} |}
\hline
Algorithm name & Format & Library source & Mod.\\ \hline
Gauss-Seidel solver &  CSR &  Intel MKL~\cite{wang2014intel} & None \\ \hline
Gauss-Seidel solver &  BCSR &  Intel MKL~\cite{wang2014intel} & None \\ \hline
Incomplete LU &  CSR &  Intel MKL~\cite{wang2014intel} & None \\ \hline
Incomplete Cholesky & CSC and CSR &  SparseLib++~\cite{pozo1996sparselib++} & None \\ \hline
Forward solve & CSC & Sympiler~\cite{cheshmi2017sympiler} & None \\  \hline
Forward solve & CSR & \cite{vuduc2002automatic} & None \\  \hline
Sparse MV Multiply  & CSR & Common & None \\  \hline
Static Left Cholesky &  CSC &  Sympiler~\cite{cheshmi2017sympiler} & P\tnote{a}\, + R\tnote{b} \\ \hline
\end{tabular}%
\label{tab:suite}
\begin{tablenotes}\footnotesize
\item [a] Privatization of temporary arrays
\item [b] Removal of dynamic index array updates
\end{tablenotes}
\end{threeparttable}
\end{center}
\end{table}

We modified onr of the benchmarks, left Cholesky, to make temporary arrays
privatizable and to remove dynamic index array updates so that the compiler can
analyze the sparse code.

\textbf{Left Cholesky}:  
This code has following changes compared to 
a more common implementation in CSparse~\cite{davis2006direct}:
\textit{(i) Privatization of temporary arrays:} We analyzed dependences between 
reads and writes to temporary arrays to detect privatizable arrays.  This
can be challenging for a compiler to do with  
sparse codes since accesses to these arrays are irregular.
We set the values of these arrays 
to zero at the beginning of each loop so 
a compiler could identify them as privatizable.
\textit{(ii) Removal of dynamic index array updates: }
Previous data dependence analysis work 
focuses on cases where index arrays are not updated.
However, in some numerical codes, updating 
index arrays is a common optimization.
We refer to this as \emph{dynamic index array updates},
and it usually occurs 
when the nonzero structure of an output matrix
is modified in the sparse code during the computation. 
This would make 
dependence analysis very complicated for the compiler.
We removed dynamic index arrays by partially decoupling symbolic
analysis from the numerical code in these benchmarks.
\emph{Symbolic analysis} here refers to terminology used in the numerical
computing community. Symbolic analysis uses the nonzero pattern of the matrix
to determine computation patterns and the sparsity of the resulting data.
To remove dynamic index array updates, we decouple symbolic analysis from
the code similar to the approach used by
~\cite{cheshmi2017sympiler}.

{\bf Performance Impact:}
The changes made to Left Cholesky do not have a 
noticeable effect on the code performance. Based on our 
experiments using five matrices\footnote{Problem1, rdb450l,
wang2, ex29, Chebyshev2} from the Florida Sparse 
Matrix Collection~\cite{davis2011university}  
the performance cost of these modifications is on average less 
than 10\% than the original code.

\subsection{Relevance of Index Array Properties}
We have extracted the constraints to test for dependences that are carried by
the outermost loop for the sparse matrix codes in Table~\ref{tab:suite}. A
total of 124 data dependences relations were collected from the benchmarks.
Of those 124, only 83 of them were unique, the repetition coming from accesses with 
same access indices in the same statements, or other situations.
Table~\ref{tab:diffInfo} summarizes the index array assertions relevant to the
benchmarks.

\begin{table}
\begin{center}
\caption{Categorization of index array properties in our evaluation of their
utility in detecting unsatisfiability.}
\label{table:properties}
\begin{tabular}{| p{1.25in} | p{2.5in} | p{1.25in} |}
\hline
\makecell{Array property} & Formulation with examples from Left Cholesky code 
& What codes found in \\ \hline
Monotonocity &
($ x_1 < x_2 \Leftrightarrow lcolptr(x_1) < lcolptr(x_2)$). & All \\ \hline
Correlated & 
($x_1 = x_2 \Rightarrow rowPtr(x_1) \le diagPtr(x_2) $). & Incomplete LU0, \\
Monotonicity & ($ x_1 < x_2 \Rightarrow diagPtr(x_2) < rowPtr(x_1)$). & 
 \\ \hline
Triangular & ($lcolptr(x_1) < x_2 \Rightarrow x_1 < lrow(x_2) $). & Cholesky's,\\
Matrix & ($x_1 < prunePtr(x_2) \Rightarrow pruneSet(x_1) < x_2 $). & Forward Solves \\
\hline
\end{tabular}%
\label{tab:diffInfo}
\end{center}
\vspace*{-.2in}
\end{table}

We are not claiming to have found all the array 
properties that exist either in our example suite nor in general. 
Also, we only consider dependence relations
for outermost loops, however, dependence relations can be extracted for other 
loop levels in a loop nest and can be used 
for vectorization 
and in other applications of dependence analysis.

\begin{figure}[ht]
\centering
\includegraphics[width=.75\columnwidth]{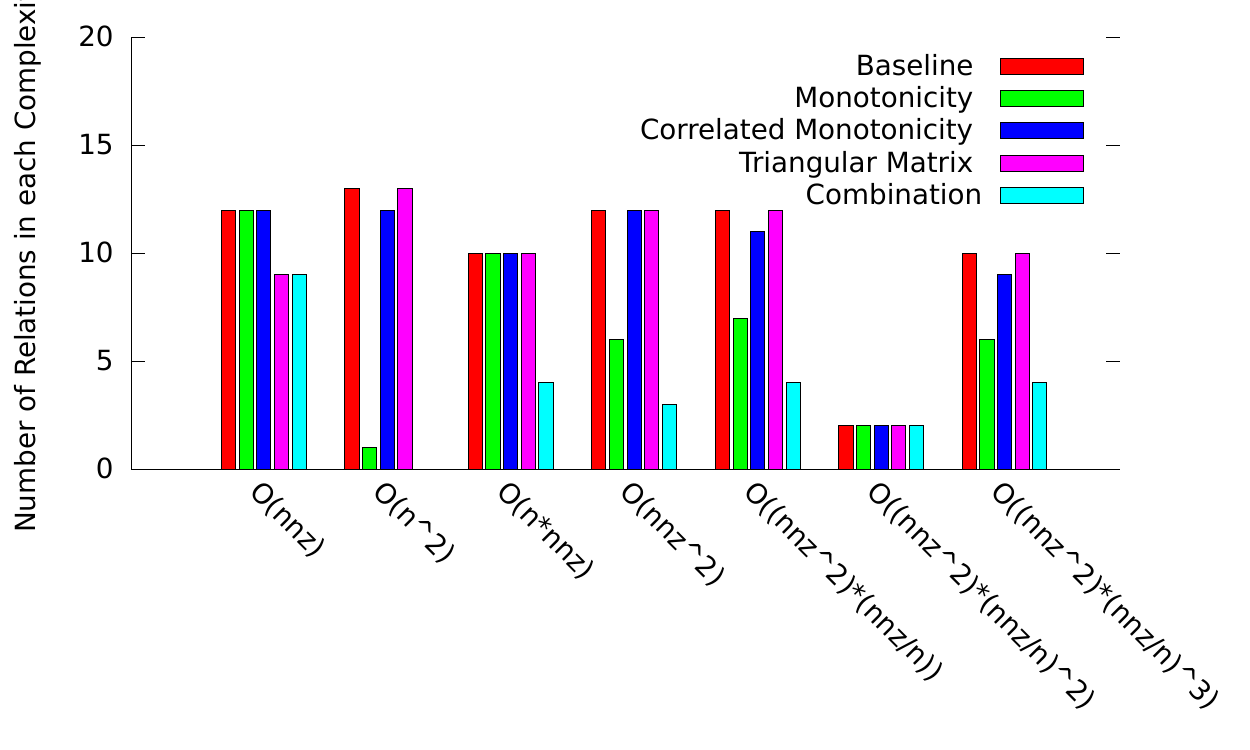}
\caption{Reduction in the number of different inspectors' complexities after adding
array properties individually and in combination.
Please note, $nnz$ is number of non-zeros, 
and $n$ is number of
columns or rows in a matrix.
The array properties discussed in the paper can help us detect 45  
relations as unsatisfiable out of 71 baseline relations.  
Note, the number of unsatisfiable relations detected with 
combination of information is not the accumulation
of all others. Sometimes combination of information
together helps detect unsatisfiability.}
\label{fig:indCom}
\end{figure}

\subsection{Detecting Unsatisfiability}
\label{detectUnsat}
In this section, we show the impact of using index array properties
to detect unsatisfiability for the relations collected 
from dependences from our benchmark suite.
To not conflate the impact of the index
array properties that we are evaluating with what 
traditional methods are capable of, we first apply functional consistency 
in the theory of Presburger arithmetic combined with 
uninterpreted functions ~\cite{Shostak79}.
This detects 12 dependences as unsatisfiable. 
Nevertheless, we must note that, all of the 12 dependences have inconsistencies in 
their affine parts and functional consistency does not help detect 
any more unsatisfiable relations; like the first two 
dependences from the Forward Solve CSR example in
Section~\ref{sec:wavefront}. After detecting 12 out of 83
dependences as unsatisfiable we are left with 71 dependences to use in our
evaluation. We call these 71 dependences that are satsfiable just by looking 
at their affine constraints our baseline.

Figure~\ref{fig:indCom} categorizes the complexity of an
inspector for each dependence into 7 different classes in total. 
In this figure, $nnz$ is number of non-zeros, and for simplicity
$n$ denotes the number of rows or columns of the input matrix.
The black bar, ``baseline'', in each class
shows the baseline number of relations with that complexity in our suite.
The  bars show how many dependences would remain
after we instantiate certain index array properties. 
The last bar in each class, the red bar, 
shows the effect of adding all the information in combination.

The main observations from analyzing Figure~\ref{fig:indCom} are as
follows:
(1) Combining the array properties and non-domain information has 
the biggest impact and helps detect significantly more unsatisfiable 
dependences than any single property.
Combining all the index array properties helped us detect 45 out of 71 relations 
as unsatisfiable, with 26 remaining as maybe satisfiable.
(2)  Monotonicity has the highest impact on detecting unsatisfiable relations
when array properties are applied independently. 
(3) The Triangular Matrix property helped detect 3 relations 
when applied independently and 11 more in combination 
with Monotonicity (not obvious in the figure). 
This property helped us detect unsatisfiability in cases 
where Monotonicity was completely handicapped; see the 
$nnz$ and $nnz*n$ classes in Figure~\ref{fig:indCom}.

\subsection{Simplifying Inspector Complexity Utilizing Equalities}
\label{sec:evalEq}
As stated in the previous section, 
instantiating index array properties results in
45 out of 71 dependence relations being detected as unsatisfiable.
At this point, without any further simplification,
to perform a partial parallelism transformation, 
inspectors are needed for the remaining 26 dependences. 
One question we can ask about those 26 inspector is whether
their complexity is even reasonable.  We consider a runtime dependence
analysis  complexity reasonable, if it is bound by the complexity of the 
original computation.
The computations would certainly
do much more operations compared to the analysis as 
numerical algorithms
usually call these computations several time for the same sparse
matrix nonzero structure.
Thus runtime data dependence analysis is reasonable 
if it is the same complexity as the original computation. 
Nonetheless, for numerical algorithms, 
it is common to aim for a runtime data dependence analysis
that is of $O(nnz)$, 
where $nnz$ is the number of nonezeros in the input. 

By instantiating index array properties with expressions
from the data dependences, it is also possible to derive
equalities between some of the iterators in the dependence.
These new useful equalities can be used to eliminate extra loops in 
the runtime inspector. 
Table~\ref{tab:insReduction} shows that the additional equalities increases 
the number of dependence relations with reasonable complexities ($\le
\mathrm{kernel}$).  For instance, the Left Cholesky code have 4 high
complexity dependence relation left. As illustrated in
Section~\ref{sec:example}, the additional equalities can be used to reduce
the complexity of all those relations.  Finding equalities also help reduce the
complexity of 4 dependences for Incomplete Cholesky0 and 2 dependences of
Incomplete LU0 to become reasonable.

We should also mention that in addition to these 10 complexity reductions,
the complexity of another 4 dependence relations were reduced. However,
the complexity after simplification is still higher than the kernel, and hence
these simplifications are not visible in Table~\ref{tab:insReduction}.

\begin{table*}
\begin{center}
\caption{Effect of simplifications based
additional equalities (Section~\ref{sec:simplifying-equality}) and redundancy
elimination (Section~\ref{sec:simplifying-superset}) on the remaining 26 maybe
satisfiable dependences for each code in the benchmark. The Total
columns show the number of dependence relations that needs to be checked at
runtime. The $\le \mathrm{kernel}$ columns show the number of such tests that
have the same or lower complexity than the kernel.  Equality Impact is the
numbers after using additional equalities, reducing the number of high complexity checks.
Supserset Impact is the composed effect of using supserset relations after
adding equalities, reducing the total number of checks.
}
\begin{tabular}{| p {3.4 cm} || p {1.4 cm} | p {1.4 cm} || p {1.3 cm}  | p {1.3 cm} || p {1.3 cm} | p {1.3 cm} |}
\hline
Kernel name
& \multicolumn{2}{|c||}{Remaining satisfiables} 
& \multicolumn{2}{|c||}{Equality Impact} & \multicolumn{2}{|c|}{Superset Impact}
\\ \hline
  & \textcolor{green}{ $\le$ kernel } & Total 
& \textcolor{green}{ $\le$ kernel } & Total 
& \textcolor{green}{ $\le$ kernel } &  Total
\\ \hline

Gauss-Seidel CSR    &  2  & 2 &  2  &  2  &  2  &  2  \\ \hline
Gauss-Seidel BCSR   &  4  & 4 &  4  &  4  &  2  &  2  \\ \hline
Incomplete LU       &  0  & 4 &  2  &  4  &  2  &  4  \\ \hline
\textbf{Incomplete Cholesky} &  \textbf{1}  & \textbf{9} &  \textbf{5}  &  \textbf{9}  &  \textbf{2}  &  \textbf{2}  \\ \hline
Forward solve CSR   &  1  & 1 &  1  &  1  &  1  &  1  \\ \hline
Forward solve CSC   &  2  & 2 &  2  &  2  &  1  &  1  \\ \hline
Sparse MV Mul.      &  0  & 0 &  0  &  0  &  0  &  0  \\ \hline
\textbf{Left Cholesky}       &  \textbf{0}  & \textbf{4} &  \textbf{4}  &  \textbf{4}  &  \textbf{2}  &  \textbf{2}  \\ \hline
\end{tabular}%
\label{tab:insReduction}
\end{center}
\end{table*}

\subsection{Impact of Utilizing Superset Relationship}
\label{eval:sup}
The superset relations we identify uncovers dependence relations that are
redundant. We can discard the dependence relations that are found to be
subsets of another and only generate runtime inspectors for remaining
relations. As shown in Table~\ref{tab:insReduction}, this results in 
fewer dependence relations to be checked at runtime. Most notably, the 
number of runtime checks were reduced from 4 to 2 for Left Cholesky, 
and both of those dependences are less comlex than the original algorithm.

As discussed in Section~\ref{sec:simplifying-superset}, the superset relation
may reveal that a relation is redundant being subset of another relation with 
lower complexity. The Incomplete Cholesky kernel were left with 4 
expensive relations even after adding equalities. 
As you can see in Table~\ref{tab:insReduction}, these relations are removed 
from runtime checks by identifying the superset relations. 
For Incomplete Cholesky kernel, we have found 2 relations with less than 
orginal algorithm complexity to be superset of all the dependences that
we need to have a runtime check for. The composed effect of our proposed
technique reduces the inspector cost to 2 or fewer inexpensive tests for all of
our kernels, except for the Incomplete LU.

\subsection{Putting It All Together}
We have presented a series of techniques to simplify dependence relations with
the main motivation being automatic generation of efficient inspector code.
Our approach aims to simplify the dependence relations starting from array
properties that can be succinctly specified by the experts. We show that the
array properties can be used to automatically disprove a large number of
potential dependences, as well as reduce the complexity of remaining
dependences. Combined with a method for detecting redundancies in dependence
tests, we are able to generate efficient inspectors.

Table~\ref{tab:suiteIns} summarizes the impact of our proposed 
approach on inspector complexity. It is interesting to note that 
Incomplete LU0 is the only kernel left with expensive inspector 
(more complex than kernel). This case is discussed further in Section~\ref{sec:limit}.

\begin{table*}
\begin{center}
\caption{The impact of our simplifications on inspector complexity.
The baseline inspector complexity is when all possible dependences are tested at runtime, without
using any of the simplifications proposed in this paper. The simplified
inspector complexity reports the final cost of inspection generated by our approach.
The overall complexity of inspectors decreases considerably.
The complexity of the kernels are included for comparison; k and K denote
constant factors, with K signaling a bigger number.}
\begin{tabular}{| p {3 cm} | p {3.7 cm} | p {3 cm} || p {2.7 cm} |}
\hline
Kernel name & Inspector complexity & Simplified inspector & Kernel complexity\\ \hline
Gauss-Seidel CSR & $(n) + 2 (nnz)$  &  $2 (nnz)$ & $k (nnz)$  \\ \hline
Gauss-Seidel BCSR & $4 (n) + 4 (nnz)$  &  $2 (nnz)$ & $k (nnz)$ \\ \hline
Incomplete LU CSR 
& $4 (nnz \times (nnz/n)) + (n^2) + 2 (n \times nnz) + 2 (nnz^2) + 2 (nnz^2 \times (nnz/n)^2) + 2 (nnz^2  \times (nnz/n)^3)$
& $2 (nnz  \times (nnz/n)^2) + 2 (nnz \times (nnz/n)^4)$ & $K (nnz  \times (nnz/n)^2)$ \\ \hline
Incomplete Cholesky CSR 
&  $10 (n^2) + 8 (nnz^2) + 6 (nnz^2  \times (nnz/n)) + 4 (nnz^2  \times (nnz/n)^3)$ 
&  $(nnz  \times (nnz/n)) + (nnz  \times (nnz/n)^2) $  & $K (nnz  \times (nnz/n)^2)$ \\\hline
Forward solve CSC & $3 (n) + 4 (nnz)$ & $nnz$ &  $k (nnz)$ \\  \hline
Forward solve CSR & $(n) + 2 (nnz) $ & $nnz$ &  $k (nnz)$ \\ \hline
Sparse MV Mul. CSR & $3 (n)$ & $0$ & $k (nnz \times (nnz/n))$ \\  \hline
Left Cholesky CSC & $8 (n \times nnz) + 4 (n^2)$  & $2(nnz)$  & $K (nnz \times (nnz/n))$ \\ \hline
\end{tabular}%
\label{tab:suiteIns}
\end{center}
\end{table*}

\subsection{Discussion: Limitations}
\label{sec:limit}
Table~\ref{tab:suiteIns} demonstrates that
our method significantly reduces both the number of runtime checks and their complexity. 
Nonetheless, our approach is not free of limitations, which are discussed in this section.

Two of the original kernels include dynamic index arrays and temporary arrays that
require privatization. As discussed in Section~\ref{sec:benchmark},
these kernels can be preprocessed such that it can be accepted by our compiler. This preprocessing
is currently done manually.

Using the associativity of reductions is important for Forward Solve CSC
and Incomplete Cholesky0.  We do not automate the reduction detection in this
paper, as it is a complex task on its own.  It is common for compilers and
programming models, such as openMP, to provide pragma interfaces for
programmers to signal which update should be considered as a reduction. We have
followed the same approach.

Incomplete LU0 has two dependence relations that has higher complexity
than the kernel, even with domain information.  Related work by~\cite{Venkat:2016} presents
approximation techniques that reduce the inspector complexity for these high
complexity relation to $nnz \times (nnz/n)$.  Such approximation can
potentially result in loss of some parallelism.  Nevertheless,
~\cite{Venkat:2016} show that the approximation of dependences does not
significantly affect the performance of the partial parallelism for this code.
We have not used approximations in our work, but it would be interesting to see
how the two approaches can be combined.

\section{Related Work}

Array data dependence analysis has been used for a variety of applications,
including automatic parallelization~\cite{Paek2002}, locality optimization~\cite{Wolfe:1989},
communication generation, program slicing~\cite{Pugh97}, detecting race
conditions~\cite{Zheng:2015},
and high-level synthesis~\cite{Alle2013}.
For sparse matrix codes, this analysis is made more difficult
due to indirection through index arrays, such that the source and sink
of dependences cannot be resolved until their values are available at runtime.
For these and other situations where dependences arise that cannot
be resolved until runtime, a number of techniques for compile time and
runtime dependence analysis have been developed.

\subsection{User-Provided Assertions}

\cite{McKinley} exploit  user assertions about index arrays 
to increase the precision
of dependence testing. The assertions certify common
properties of index arrays, e.g., an index array can be a permutation
array, monotonically increasing, and monotonically
decreasing.
\cite{Lin:2000:CAI} present a compile time analysis
for determining index array properties, such as monotonicity.
They use the analysis results for  parallelization of sparse matrix
computations.

Our approach also uses these assertions, but in
addition we use more domain-specific assertions and provide
a way to automate the general use of such assertions.
In this paper, the idea of applying constraint instantiation of universally
quantified constraints as is done in SMT solvers to find unsatisfactory
dependences is novel and the assertions about index arrays we use
are more general.

\subsection{Proving Index Arrays Satisfy the Assertions}
In this work, we assume that the assertions provided by the programmer is
correct. It is useful to verify the user-provided assertions by analyzing the
code that constructs the sparse matrix data structures. There is a
large body of work in abstract interpretation that address this problem.

The major challenge in verifying the assertion about programs that manipulate
arrays is the trade-off between scalability and precision. When there is a
large number of updates to an array, keeping track of individual elements do
not scale, but approximating the whole array as a single summary significantly
degrades the precision. Many techniques to verify/infer important properties
about array contents from programs have been developed, e.g.,
\cite{cousot2011parametric,gopan2005framework,halbwachs2008}.

In the work by \cite{kovacs-et-al-vmcai-2010}, the authors present an approach
for inferring shape invariants for matrices.
While their work does not deal with sparse matrices and index arrays,
it may help generate domain-specific assertions that we could
employ to show that the data dependences are unsatisfiable.

The main subject of our work - dependence tests - does not involve array
updates, since all the index arrays, which alter the control-flow and indexing
of the data arrays, are not updated. This makes the verification of the
assertions a closely related but orthogonal problem, which we do not address in
this paper.

\subsection{More General Quantifier Elimination Techniques}

The area of SMT-solving is advancing at a significant pace; the webpage for
SMT-COMP\footnote{\url{http://smtcomp.sourceforge.net/2017/}} provides a list
of virtually all actively developed solvers, and how they fared in each theory
category.
As these solvers are moving into a variety of domains,
quantifier instantiation and elimination has become a topic of 
central interest.
Some of the recent work in this area are: E-matching~\cite{moura2007},
Model-Based~\cite{Ge2009}, Conflict-Based~\cite{reynolds2014finding}, and
Counter-Example Guided~\cite{reynolds2015counterexample}.

These efforts make it clear that
quantifier instantiation is challenging, and is an area of active development.
SMT solvers often rely on 
heuristic-driven instantiations to show unsat for difficult problems.
In this context,
our work can be viewed as heuristic instantiation where 
the heuristic is inspired by decidable fragments of the array theory.

Dependence constraints with universally quantified assertions
are related to the first order theory fragments
described by~\cite{Bradley2006} as undecidable extensions
to their array theory fragment.
However, ~\cite{Loding2017} claim that the proofs for undecidability
of extension theories by~\cite{Bradley2006} are incorrect,
and declare their decidability status as an open problem.
Regardless of whether the theory fragment that encompasses our
dependence constraints is decidable or not following is true: if we
soundly prove that a relation is unsatisfiable just with compile time
information, the unsatisfiability applies in general, and having
runtime information would not change anything. However, if a dependence
detected to be satisfiable just with compile time information, we need
to have runtime tests to see if it is actually satisfiable given runtime
information, and even if it is, run time tests would determine
for what values the dependence holds.

\subsection{Dependence Analysis for Full Parallelization}

Some compilation techniques have been developed to extend the dependence analysis to
sparse, or non-affine programs~\cite{Poly2010}. These
techniques extend to non-affine programs of various forms: while loops,
polynomial expressions, function calls, data-dependent conditions, and
indirection arrays. The outcome of such analysis is often an approximation,
which is quite pessimistic for sparse computations involving indirection
arrays. The focus of our work is not to identify (approximated) dependences,
but to reduce the cost of runtime dependence checks by disproving potential
dependences as much as possible at compile-time.

The work by \cite{pugh98constraintbased} also formulate the
problem in the theory of Presburger sets with uninterpreted functions.
However, they only allow affine expressions of unquantified variables as
indexing expressions to the function symbols, excluding some of the examples in
this paper.  They propose an analysis to identify conditions for a dependence
to exist through the use of gist operator that simplifies the constraint system
given its context. The result of this analysis may involve uninterpreted
functions, and can be used to query the programmer for their domain knowledge.
This is an interesting direction of interaction that complements our work.

Several runtime approaches focus on identifying loops
we denoted \textit{fully parallel} whose iterations are
independent and can safely execute in 
parallel~\cite{barthou97fuzzy,pugh98constraintbased,Moon:1999}
or speculatively execute in parallel while testing safety~\cite{Rauchwerger:1999}.

\subsection{Dependence Analysis for Wavefront Parallelization}

For sparse codes, even when loops carry dependences, the dependences 
themselves may be sparse, and it may be possible to execute some iterations of the
loop in parallel (previously denoted \textit{partially parallel}.
The parallelism is captured in a task graph, and typically executed as a parallel wavefront.
A number of prior works write specialized code to derive this task graph as part 
of their application~\cite{Saltz91,Rauchwerger95,Zhuang09,Bell:SpMV:SC:2009,Park2014,Jongsoo14SC}
or with kernel-specific code generators~\cite{pOski12}.
For example, Saltz and Rothbergs worked on manual parallelization of  
sparse triangular solver codes  in the 1990s~\cite{Saltz90,Rothberg92}.
There is also more recent work on optimizing sparse triangular solver
NVIDIA GPUs and Intel's multi-core CPUs~\cite{rennich2016accelerating, wang2014intel}.
Even though these manual optimizations have been successful
at achieving high performance in some cases, significant programmer
effort has to be invested for each of these codes and automating
these parallelization strategies can significantly reduce this effort.

Other approaches automate the generation of inspectors that
find task-graph, wavefront or partial parallelism. \cite{rauchwerger95scalable}
and others~\cite{HuangJBJA13} have developed efficient and parallel inspectors
that maintain lists of iterations that read and write each memory location.
By increasing the number of dependences found unsatisfiable,
the approach presented in this paper reduces
the number of memory accesses that would need to be tracked.
For satisfiable dependences, there is a tradeoff between inspecting
iteration space dependences versus maintaining data for each memory
access. That choice could be made at runtime. There are also
other approaches for automatic generation of inspectors that
have looked at simplifying the inspector by finding equalities,
using approximation, parallelizing the inspector, and applying
point-to-point synchronization to the executor~\cite{Venkat:2016}.

\subsection{Algorithm-Specific Data Dependence Analysis}

An algorithm-specific approach to represent data dependences and optimize
memory usage of sparse factorization algorithms such as Cholesky
~\cite{pothen2004elimination} uses an \textit{elimination tree}, but to the
best of our knowledge, this structure is not derived automatically from source code.
When factorizing a column of a sparse
matrix, in addition to nonzero elements of the input matrix new nonzero
elements, called fill-in, might be created. Since the sparse matrices
are compressed for efficiency, the additional fills during
factorization make memory allocation ahead of factorization difficult.
The elimination tree is used to predict the sparsity pattern of the L
factor ahead of factorization so the size of the factor can be computed~\cite{coleman1986predicting} or predicted~\cite{gilbert1994predicting,gilbert1993predicting},
and captures a potential parallel schedule of the tasks.
Prior work has investigated the applicability of the elimination
tree for dependence analysis for  parallel implementation
~\cite{george1989communication,gilbert1992highly,pothen1993mapping,                                                                                                                                              
karypis1995high,schenk2002two,henon2002pastix,hogg2010design,                                                                                                                                                    
zheng2015gpu,rennich2016accelerating}.
Some techniques such 
as~\cite{pothen1993mapping,george1989communication,henon2002pastix} 
use the elimination tree for static scheduling while
others use it for runtime scheduling.

\section{Conclusion}

In this paper, we present an automated approach for 
showing sparse code data dependences are unsatisfiable or if
not reducing the complexity for later runtime analysis.
Refuting a data dependence brings benefits to many areas of sparse matrix code
analysis, including verification and loop optimizations such as parallelization, pipelining,
or tiling by
completely eliminating the high runtime costs of
deploying runtime dependence checking.
Additionally, when a dependence remains satisfiable, our 
approach of performing constraint instantiation within
the context of the Integer Set Library (ISL)
enables equalities and subset relationships to be 
derived that simplify the runtime complexity
of inspectors for a case study with wavefront parallelism.
Parallelization of these sparse numerical methods is 
an active research area today, but one where most current approaches
require {\em manual parallelization}.
It is also worth noting that without inspector complexity reduction, most
inspectors would timeout, thus underscoring the pivotal role of  the work in
this paper in enabling parallelization and optimization of sparse codes.
Our results are established over
71 dependences extracted from 8 sparse numerical methods.


\bibliography{sparseDependenceAnalysis_arxiv}


\begin{thebibliography}{74}


\ifx \showCODEN    \undefined \def \showCODEN     #1{\unskip}     \fi
\ifx \showDOI      \undefined \def \showDOI       #1{#1}\fi
\ifx \showISBNx    \undefined \def \showISBNx     #1{\unskip}     \fi
\ifx \showISBNxiii \undefined \def \showISBNxiii  #1{\unskip}     \fi
\ifx \showISSN     \undefined \def \showISSN      #1{\unskip}     \fi
\ifx \showLCCN     \undefined \def \showLCCN      #1{\unskip}     \fi
\ifx \shownote     \undefined \def \shownote      #1{#1}          \fi
\ifx \showarticletitle \undefined \def \showarticletitle #1{#1}   \fi
\ifx \showURL      \undefined \def \showURL       {\relax}        \fi
\providecommand\bibfield[2]{#2}
\providecommand\bibinfo[2]{#2}
\providecommand\natexlab[1]{#1}
\providecommand\showeprint[2][]{arXiv:#2}

\bibitem[\protect\citeauthoryear{??}{chi}{2018}]%
        {chill:ctop}
 \bibinfo{year}{2018}\natexlab{}.
\newblock \bibinfo{title}{CTOP research group webpage at Utah}.
\newblock   (\bibinfo{year}{2018}).
\newblock
\urldef\tempurl%
\url{http://ctop.cs.utah.edu/ctop/?page_id=21}
\showURL{%
\tempurl}


\bibitem[\protect\citeauthoryear{Alle, Morvan, and Derrien}{Alle
  et~al\mbox{.}}{2013}]%
        {Alle2013}
\bibfield{author}{\bibinfo{person}{Mythri Alle}, \bibinfo{person}{Antoine
  Morvan}, {and} \bibinfo{person}{Steven Derrien}.}
  \bibinfo{year}{2013}\natexlab{}.
\newblock \showarticletitle{Runtime Dependency Analysis for Loop Pipelining in
  High-level Synthesis}. In \bibinfo{booktitle}{\emph{Proceedings of the 50th
  Annual Design Automation Conference}} \emph{(\bibinfo{series}{DAC '13})}.
  \bibinfo{publisher}{ACM}, \bibinfo{address}{New York, NY, USA}, Article
  \bibinfo{articleno}{51}, \bibinfo{numpages}{10}~pages.
\newblock
\showISBNx{978-1-4503-2071-9}
\urldef\tempurl%
\url{https://doi.org/10.1145/2463209.2488796}
\showDOI{\tempurl}


\bibitem[\protect\citeauthoryear{Atzeni, Gopalakrishnan, Rakamaric, Ahn,
  Laguna, Schulz, Lee, Protze, and M{\"{u}}ller}{Atzeni et~al\mbox{.}}{2016a}]%
        {AtzeniGRALSLPM16}
\bibfield{author}{\bibinfo{person}{Simone Atzeni}, \bibinfo{person}{Ganesh
  Gopalakrishnan}, \bibinfo{person}{Zvonimir Rakamaric},
  \bibinfo{person}{Dong~H. Ahn}, \bibinfo{person}{Ignacio Laguna},
  \bibinfo{person}{Martin Schulz}, \bibinfo{person}{Gregory~L. Lee},
  \bibinfo{person}{Joachim Protze}, {and} \bibinfo{person}{Matthias~S.
  M{\"{u}}ller}.} \bibinfo{year}{2016}\natexlab{a}.
\newblock \showarticletitle{{ARCHER:} Effectively Spotting Data Races in Large
  OpenMP Applications}. In \bibinfo{booktitle}{\emph{2016 {IEEE} International
  Parallel and Distributed Processing Symposium, {IPDPS} 2016, Chicago, IL,
  USA, May 23-27, 2016}}. \bibinfo{pages}{53--62}.
\newblock
\urldef\tempurl%
\url{https://doi.org/10.1109/IPDPS.2016.68}
\showDOI{\tempurl}


\bibitem[\protect\citeauthoryear{Atzeni, Gopalakrishnan, Rakamaric, Ahn,
  Laguna, Schulz, Lee, Protze, and Müller}{Atzeni et~al\mbox{.}}{2016b}]%
        {Atzeni2016}
\bibfield{author}{\bibinfo{person}{S. Atzeni}, \bibinfo{person}{G.
  Gopalakrishnan}, \bibinfo{person}{Z. Rakamaric}, \bibinfo{person}{D.~H. Ahn},
  \bibinfo{person}{I. Laguna}, \bibinfo{person}{M. Schulz},
  \bibinfo{person}{G.~L. Lee}, \bibinfo{person}{J. Protze}, {and}
  \bibinfo{person}{M.~S. Müller}.} \bibinfo{year}{2016}\natexlab{b}.
\newblock \showarticletitle{ARCHER: Effectively Spotting Data Races in Large
  OpenMP Applications}. In \bibinfo{booktitle}{\emph{2016 IEEE International
  Parallel and Distributed Processing Symposium (IPDPS)}}.
  \bibinfo{pages}{53--62}.
\newblock
\showISSN{1530-2075}
\urldef\tempurl%
\url{https://doi.org/10.1109/IPDPS.2016.68}
\showDOI{\tempurl}


\bibitem[\protect\citeauthoryear{Banerjee, Gelernter, Nicolau, and
  Padua}{Banerjee et~al\mbox{.}}{1993}]%
        {PW93}
\bibfield{editor}{\bibinfo{person}{Uptal Banerjee}, \bibinfo{person}{David
  Gelernter}, \bibinfo{person}{Alex Nicolau}, {and} \bibinfo{person}{David
  Padua}} (Eds.). \bibinfo{year}{1993}\natexlab{}.
\newblock \bibinfo{booktitle}{\emph{An Exact Method for Analysis of Value-based
  Array Data Dependences}}. \bibinfo{publisher}{Springer-Verlag},
  \bibinfo{address}{London, UK}.
\newblock


\bibitem[\protect\citeauthoryear{Barthou, Collard, and Feautrier}{Barthou
  et~al\mbox{.}}{1997}]%
        {barthou97fuzzy}
\bibfield{author}{\bibinfo{person}{Denis Barthou},
  \bibinfo{person}{Jean-Fran{\c{c}}ois Collard}, {and} \bibinfo{person}{Paul
  Feautrier}.} \bibinfo{year}{1997}\natexlab{}.
\newblock \showarticletitle{Fuzzy Array Dataflow Analysis}.
\newblock \bibinfo{journal}{\emph{J. Parallel and Distrib. Comput.}}
  \bibinfo{volume}{40}, \bibinfo{number}{2} (\bibinfo{year}{1997}),
  \bibinfo{pages}{210--226}.
\newblock


\bibitem[\protect\citeauthoryear{Basumallik and Eigenmann}{Basumallik and
  Eigenmann}{2006}]%
        {Basumallik06}
\bibfield{author}{\bibinfo{person}{Ayon Basumallik} {and}
  \bibinfo{person}{Rudolf Eigenmann}.} \bibinfo{year}{2006}\natexlab{}.
\newblock \showarticletitle{Optimizing irregular shared-memory applications for
  distributed-memory systems}. In \bibinfo{booktitle}{\emph{Proceedings of the
  Eleventh ACM SIGPLAN Symposium on Principles and Practice of Parallel
  Programming}}. \bibinfo{publisher}{ACM Press}, \bibinfo{address}{New York,
  NY, USA}, \bibinfo{pages}{119--128}.
\newblock


\bibitem[\protect\citeauthoryear{Bell and Garland}{Bell and Garland}{2009}]%
        {Bell:SpMV:SC:2009}
\bibfield{author}{\bibinfo{person}{Nathan Bell} {and} \bibinfo{person}{Michael
  Garland}.} \bibinfo{year}{2009}\natexlab{}.
\newblock \showarticletitle{Implementing sparse matrix-vector multiplication on
  throughput-oriented processors}. In \bibinfo{booktitle}{\emph{SC '09:
  Proceedings of the Conference on High Performance Computing Networking,
  Storage and Analysis}}. \bibinfo{publisher}{ACM}, \bibinfo{address}{New York,
  NY, USA}, \bibinfo{pages}{1--11}.
\newblock


\bibitem[\protect\citeauthoryear{Benabderrahmane, Pouchet, Cohen, and
  Bastoul}{Benabderrahmane et~al\mbox{.}}{2010}]%
        {Poly2010}
\bibfield{author}{\bibinfo{person}{Mohamed-Walid Benabderrahmane},
  \bibinfo{person}{Louis-No{\"e}l Pouchet}, \bibinfo{person}{Albert Cohen},
  {and} \bibinfo{person}{C{\'e}dric Bastoul}.} \bibinfo{year}{2010}\natexlab{}.
\newblock \showarticletitle{The Polyhedral Model Is More Widely Applicable Than
  You Think}. In \bibinfo{booktitle}{\emph{Compiler Construction}},
  Vol.~\bibinfo{volume}{LNCS 6011}. \bibinfo{publisher}{Springer-Verlag},
  \bibinfo{address}{Berlin, Heidelberg}.
\newblock


\bibitem[\protect\citeauthoryear{Bradley, Manna, and Sipma}{Bradley
  et~al\mbox{.}}{2006}]%
        {Bradley2006}
\bibfield{author}{\bibinfo{person}{Aaron~R. Bradley}, \bibinfo{person}{Zohar
  Manna}, {and} \bibinfo{person}{Henny~B. Sipma}.}
  \bibinfo{year}{2006}\natexlab{}.
\newblock \bibinfo{booktitle}{\emph{What's Decidable About Arrays?}}
\newblock \bibinfo{publisher}{Springer Berlin Heidelberg},
  \bibinfo{address}{Berlin, Heidelberg}, \bibinfo{pages}{427--442}.
\newblock
\showISBNx{978-3-540-31622-0}
\urldef\tempurl%
\url{https://doi.org/10.1007/11609773_28}
\showDOI{\tempurl}


\bibitem[\protect\citeauthoryear{Brandes}{Brandes}{1988}]%
        {Brandes88}
\bibfield{author}{\bibinfo{person}{T. Brandes}.}
  \bibinfo{year}{1988}\natexlab{}.
\newblock \showarticletitle{The importance of direct dependences for automatic
  parallelism}. In \bibinfo{booktitle}{\emph{Proceedings of the International
  Conference on Supercomputing}}. \bibinfo{publisher}{ACM},
  \bibinfo{address}{New York, NY, USA}, \bibinfo{pages}{407--417}.
\newblock


\bibitem[\protect\citeauthoryear{Byun, Lin, Yelick, and Demmel}{Byun
  et~al\mbox{.}}{2012}]%
        {pOski12}
\bibfield{author}{\bibinfo{person}{Jong-Ho Byun}, \bibinfo{person}{Richard
  Lin}, \bibinfo{person}{Katherine~A. Yelick}, {and} \bibinfo{person}{James
  Demmel}.} \bibinfo{year}{2012}\natexlab{}.
\newblock \bibinfo{booktitle}{\emph{Autotuning Sparse Matrix-Vector
  Multiplication for Multicore}}.
\newblock \bibinfo{type}{{T}echnical {R}eport}.
  \bibinfo{institution}{UCB/EECS-2012-215}.
\newblock


\bibitem[\protect\citeauthoryear{Cheshmi, Kamil, Strout, and Dehnavi}{Cheshmi
  et~al\mbox{.}}{2017}]%
        {cheshmi2017sympiler}
\bibfield{author}{\bibinfo{person}{Kazem Cheshmi}, \bibinfo{person}{Shoaib
  Kamil}, \bibinfo{person}{Michelle~Mills Strout}, {and}
  \bibinfo{person}{Maryam~Mehri Dehnavi}.} \bibinfo{year}{2017}\natexlab{}.
\newblock \showarticletitle{Sympiler: Transforming Sparse Matrix Codes by
  Decoupling Symbolic Analysis}. In \bibinfo{booktitle}{\emph{Proceedings of
  the International Conference for High Performance Computing, Networking,
  Storage and Analysis}} \emph{(\bibinfo{series}{SC '17})}.
  \bibinfo{publisher}{ACM}, \bibinfo{address}{New York, NY, USA}, Article
  \bibinfo{articleno}{13}, \bibinfo{numpages}{13}~pages.
\newblock
\showISBNx{978-1-4503-5114-0}
\urldef\tempurl%
\url{https://doi.org/10.1145/3126908.3126936}
\showDOI{\tempurl}


\bibitem[\protect\citeauthoryear{Coleman, Edenbrandt, and Gilbert}{Coleman
  et~al\mbox{.}}{1986}]%
        {coleman1986predicting}
\bibfield{author}{\bibinfo{person}{Thomas~F Coleman}, \bibinfo{person}{Anders
  Edenbrandt}, {and} \bibinfo{person}{John~R Gilbert}.}
  \bibinfo{year}{1986}\natexlab{}.
\newblock \showarticletitle{Predicting fill for sparse orthogonal
  factorization}.
\newblock \bibinfo{journal}{\emph{Journal of the ACM (JACM)}}
  \bibinfo{volume}{33}, \bibinfo{number}{3} (\bibinfo{year}{1986}),
  \bibinfo{pages}{517--532}.
\newblock


\bibitem[\protect\citeauthoryear{Cousot, Cousot, and Logozzo}{Cousot
  et~al\mbox{.}}{2011}]%
        {cousot2011parametric}
\bibfield{author}{\bibinfo{person}{Patrick Cousot}, \bibinfo{person}{Radhia
  Cousot}, {and} \bibinfo{person}{Francesco Logozzo}.}
  \bibinfo{year}{2011}\natexlab{}.
\newblock \showarticletitle{A Parametric Segmentation Functor for Fully
  Automatic and Scalable Array Content Analysis}. In
  \bibinfo{booktitle}{\emph{Proceedings of the 38th Symposium on Principles of
  Programming Languages}} \emph{(\bibinfo{series}{POPL '11})}.
  \bibinfo{pages}{105--118}.
\newblock
\urldef\tempurl%
\url{https://doi.org/10.1145/1926385.1926399}
\showDOI{\tempurl}


\bibitem[\protect\citeauthoryear{Davis}{Davis}{2006}]%
        {davis2006direct}
\bibfield{author}{\bibinfo{person}{Timothy~A Davis}.}
  \bibinfo{year}{2006}\natexlab{}.
\newblock \bibinfo{booktitle}{\emph{Direct methods for sparse linear systems}}.
  Vol.~\bibinfo{volume}{2}.
\newblock \bibinfo{publisher}{Siam}.
\newblock


\bibitem[\protect\citeauthoryear{Davis and Hu}{Davis and Hu}{2011}]%
        {davis2011university}
\bibfield{author}{\bibinfo{person}{Timothy~A Davis} {and}
  \bibinfo{person}{Yifan Hu}.} \bibinfo{year}{2011}\natexlab{}.
\newblock \showarticletitle{The University of Florida sparse matrix
  collection}.
\newblock \bibinfo{journal}{\emph{ACM Transactions on Mathematical Software
  (TOMS)}} \bibinfo{volume}{38}, \bibinfo{number}{1} (\bibinfo{year}{2011}),
  \bibinfo{pages}{1}.
\newblock


\bibitem[\protect\citeauthoryear{Douglas, Hu, Kowarschik, R{\"u}de, and
  Wei\ss.}{Douglas et~al\mbox{.}}{2000}]%
        {dimeEtna00}
\bibfield{author}{\bibinfo{person}{Craig~C. Douglas}, \bibinfo{person}{Jonathan
  Hu}, \bibinfo{person}{Markus Kowarschik}, \bibinfo{person}{Ulrich R{\"u}de},
  {and} \bibinfo{person}{Christian Wei\ss.}} \bibinfo{year}{2000}\natexlab{}.
\newblock \showarticletitle{{C}ache {O}ptimization for {S}tructured and
  {U}nstructured {G}rid {M}ultigrid}.
\newblock \bibinfo{journal}{\emph{Electronic Transaction on Numerical
  Analysis}} (\bibinfo{date}{February} \bibinfo{year}{2000}),
  \bibinfo{pages}{21--40}.
\newblock


\bibitem[\protect\citeauthoryear{Feautrier}{Feautrier}{1991}]%
        {Feau91}
\bibfield{author}{\bibinfo{person}{Paul Feautrier}.}
  \bibinfo{year}{1991}\natexlab{}.
\newblock \showarticletitle{Dataflow Analysis of Array and Scalar References}.
\newblock \bibinfo{journal}{\emph{International Journal of Parallel
  Programming}} \bibinfo{volume}{20}, \bibinfo{number}{1}
  (\bibinfo{date}{February} \bibinfo{year}{1991}), \bibinfo{pages}{23--53}.
\newblock


\bibitem[\protect\citeauthoryear{Ge and de~Moura}{Ge and de~Moura}{2009}]%
        {Ge2009}
\bibfield{author}{\bibinfo{person}{Yeting Ge} {and} \bibinfo{person}{Leonardo
  de Moura}.} \bibinfo{year}{2009}\natexlab{}.
\newblock \bibinfo{booktitle}{\emph{Complete Instantiation for Quantified
  Formulas in Satisfiabiliby Modulo Theories}}.
\newblock \bibinfo{publisher}{Springer Berlin Heidelberg},
  \bibinfo{address}{Berlin, Heidelberg}, \bibinfo{pages}{306--320}.
\newblock


\bibitem[\protect\citeauthoryear{George, Liu, and Ng}{George
  et~al\mbox{.}}{1989}]%
        {george1989communication}
\bibfield{author}{\bibinfo{person}{Alan George}, \bibinfo{person}{Joseph~WH
  Liu}, {and} \bibinfo{person}{Esmond Ng}.} \bibinfo{year}{1989}\natexlab{}.
\newblock \showarticletitle{Communication results for parallel sparse Cholesky
  factorization on a hypercube}.
\newblock \bibinfo{journal}{\emph{Parallel Comput.}} \bibinfo{volume}{10},
  \bibinfo{number}{3} (\bibinfo{year}{1989}), \bibinfo{pages}{287--298}.
\newblock


\bibitem[\protect\citeauthoryear{Gilbert}{Gilbert}{1994}]%
        {gilbert1994predicting}
\bibfield{author}{\bibinfo{person}{John~R Gilbert}.}
  \bibinfo{year}{1994}\natexlab{}.
\newblock \showarticletitle{Predicting structure in sparse matrix
  computations}.
\newblock \bibinfo{journal}{\emph{SIAM J. Matrix Anal. Appl.}}
  \bibinfo{volume}{15}, \bibinfo{number}{1} (\bibinfo{year}{1994}),
  \bibinfo{pages}{62--79}.
\newblock


\bibitem[\protect\citeauthoryear{Gilbert and Ng}{Gilbert and Ng}{1993}]%
        {gilbert1993predicting}
\bibfield{author}{\bibinfo{person}{John~R Gilbert} {and}
  \bibinfo{person}{Esmond~G Ng}.} \bibinfo{year}{1993}\natexlab{}.
\newblock \showarticletitle{Predicting structure in nonsymmetric sparse matrix
  factorizations}.
\newblock In \bibinfo{booktitle}{\emph{Graph theory and sparse matrix
  computation}}. \bibinfo{publisher}{Springer}, \bibinfo{pages}{107--139}.
\newblock


\bibitem[\protect\citeauthoryear{Gilbert and Schreiber}{Gilbert and
  Schreiber}{1992}]%
        {gilbert1992highly}
\bibfield{author}{\bibinfo{person}{John~R Gilbert} {and}
  \bibinfo{person}{Robert Schreiber}.} \bibinfo{year}{1992}\natexlab{}.
\newblock \showarticletitle{Highly parallel sparse Cholesky factorization}.
\newblock \bibinfo{journal}{\emph{SIAM J. Sci. Statist. Comput.}}
  \bibinfo{volume}{13}, \bibinfo{number}{5} (\bibinfo{year}{1992}),
  \bibinfo{pages}{1151--1172}.
\newblock


\bibitem[\protect\citeauthoryear{Gopan, Reps, and Sagiv}{Gopan
  et~al\mbox{.}}{2005}]%
        {gopan2005framework}
\bibfield{author}{\bibinfo{person}{Denis Gopan}, \bibinfo{person}{Thomas Reps},
  {and} \bibinfo{person}{Mooly Sagiv}.} \bibinfo{year}{2005}\natexlab{}.
\newblock \showarticletitle{A Framework for Numeric Analysis of Array
  Operations}. In \bibinfo{booktitle}{\emph{Proceedings of the 32nd Symposium
  on Principles of Programming Languages}} \emph{(\bibinfo{series}{POPL '05})}.
  \bibinfo{pages}{338--350}.
\newblock
\showISBNx{1-58113-830-X}
\urldef\tempurl%
\url{https://doi.org/10.1145/1040305.1040333}
\showDOI{\tempurl}


\bibitem[\protect\citeauthoryear{Habermehl, Iosif, and Vojnar}{Habermehl
  et~al\mbox{.}}{2008}]%
        {Habermehl2008}
\bibfield{author}{\bibinfo{person}{Peter Habermehl}, \bibinfo{person}{Radu
  Iosif}, {and} \bibinfo{person}{Tom\'{a}\v{s} Vojnar}.}
  \bibinfo{year}{2008}\natexlab{}.
\newblock \showarticletitle{What else is Decidable About Integer Arrays?}. In
  \bibinfo{booktitle}{\emph{Proceedings of the 11th International Conference on
  Foundations of Software Science and Computational Structures}}
  \emph{(\bibinfo{series}{FOSSACS'08/ETAPS'08})}. \bibinfo{pages}{474--489}.
\newblock
\showISBNx{3-540-78497-7, 978-3-540-78497-5}


\bibitem[\protect\citeauthoryear{Halbwachs and P{\'e}ron}{Halbwachs and
  P{\'e}ron}{2008}]%
        {halbwachs2008}
\bibfield{author}{\bibinfo{person}{Nicolas Halbwachs} {and}
  \bibinfo{person}{Mathias P{\'e}ron}.} \bibinfo{year}{2008}\natexlab{}.
\newblock \showarticletitle{Discovering Properties About Arrays in Simple
  Programs}. In \bibinfo{booktitle}{\emph{Proceedings of the 29th Conference on
  Programming Language Design and Implementation}} \emph{(\bibinfo{series}{PLDI
  '08})}. \bibinfo{pages}{339--348}.
\newblock
\showISBNx{978-1-59593-860-2}
\urldef\tempurl%
\url{https://doi.org/10.1145/1375581.1375623}
\showDOI{\tempurl}


\bibitem[\protect\citeauthoryear{H{\'e}non, Ramet, and Roman}{H{\'e}non
  et~al\mbox{.}}{2002}]%
        {henon2002pastix}
\bibfield{author}{\bibinfo{person}{Pascal H{\'e}non}, \bibinfo{person}{Pierre
  Ramet}, {and} \bibinfo{person}{Jean Roman}.} \bibinfo{year}{2002}\natexlab{}.
\newblock \showarticletitle{PASTIX: a high-performance parallel direct solver
  for sparse symmetric positive definite systems}.
\newblock \bibinfo{journal}{\emph{Parallel Comput.}} \bibinfo{volume}{28},
  \bibinfo{number}{2} (\bibinfo{year}{2002}), \bibinfo{pages}{301--321}.
\newblock


\bibitem[\protect\citeauthoryear{Henzinger, Hottelier, Kov{\'{a}}cs, and
  Voronkov}{Henzinger et~al\mbox{.}}{2010}]%
        {kovacs-et-al-vmcai-2010}
\bibfield{author}{\bibinfo{person}{Thomas~A. Henzinger},
  \bibinfo{person}{Thibaud Hottelier}, \bibinfo{person}{Laura Kov{\'{a}}cs},
  {and} \bibinfo{person}{Andrei Voronkov}.} \bibinfo{year}{2010}\natexlab{}.
\newblock \showarticletitle{Invariant and Type Inference for Matrices}. In
  \bibinfo{booktitle}{\emph{Verification, Model Checking, and Abstract
  Interpretation, 11th International Conference, {VMCAI} 2010, Madrid, Spain,
  January 17-19, 2010. Proceedings}}. \bibinfo{pages}{163--179}.
\newblock


\bibitem[\protect\citeauthoryear{Hogg, Reid, and Scott}{Hogg
  et~al\mbox{.}}{2010}]%
        {hogg2010design}
\bibfield{author}{\bibinfo{person}{Jonathan~D Hogg}, \bibinfo{person}{John~K
  Reid}, {and} \bibinfo{person}{Jennifer~A Scott}.}
  \bibinfo{year}{2010}\natexlab{}.
\newblock \showarticletitle{Design of a multicore sparse Cholesky factorization
  using DAGs}.
\newblock \bibinfo{journal}{\emph{SIAM Journal on Scientific Computing}}
  \bibinfo{volume}{32}, \bibinfo{number}{6} (\bibinfo{year}{2010}),
  \bibinfo{pages}{3627--3649}.
\newblock


\bibitem[\protect\citeauthoryear{Huang, Jablin, Beard, Johnson, and
  August}{Huang et~al\mbox{.}}{2013}]%
        {HuangJBJA13}
\bibfield{author}{\bibinfo{person}{Jialu Huang}, \bibinfo{person}{Thomas~B.
  Jablin}, \bibinfo{person}{Stephen~R. Beard}, \bibinfo{person}{Nick~P.
  Johnson}, {and} \bibinfo{person}{David~I. August}.}
  \bibinfo{year}{2013}\natexlab{}.
\newblock \showarticletitle{Automatically exploiting cross-invocation
  parallelism using runtime information}. In \bibinfo{booktitle}{\emph{CGO}}.
  \bibinfo{pages}{1--11}.
\newblock


\bibitem[\protect\citeauthoryear{Karypis and Kumar}{Karypis and Kumar}{1995}]%
        {karypis1995high}
\bibfield{author}{\bibinfo{person}{George Karypis} {and} \bibinfo{person}{Vipin
  Kumar}.} \bibinfo{year}{1995}\natexlab{}.
\newblock \showarticletitle{A high performance sparse Cholesky factorization
  algorithm for scalable parallel computers}. In
  \bibinfo{booktitle}{\emph{Frontiers of Massively Parallel Computation, 1995.
  Proceedings. Frontiers' 95., Fifth Symposium on the}}. IEEE,
  \bibinfo{pages}{140--147}.
\newblock


\bibitem[\protect\citeauthoryear{Kroening and Strichman}{Kroening and
  Strichman}{2016}]%
        {Kroening2016}
\bibfield{author}{\bibinfo{person}{Daniel Kroening} {and} \bibinfo{person}{Ofer
  Strichman}.} \bibinfo{year}{2016}\natexlab{}.
\newblock \bibinfo{booktitle}{\emph{Decision Procedures: An Algorithmic Point
  of View} (\bibinfo{edition}{2nd} ed.)}.
\newblock \bibinfo{publisher}{Springer Berlin Heidelberg}.
\newblock


\bibitem[\protect\citeauthoryear{Lin and Padua}{Lin and Padua}{2000}]%
        {Lin:2000:CAI}
\bibfield{author}{\bibinfo{person}{Yuan Lin} {and} \bibinfo{person}{David
  Padua}.} \bibinfo{year}{2000}\natexlab{}.
\newblock \showarticletitle{Compiler analysis of irregular memory accesses}. In
  \bibinfo{booktitle}{\emph{Proceedings of the {ACM SIGPLAN} Conference on
  Programming Language Design and Implementation}}, Vol.~\bibinfo{volume}{35}.
  \bibinfo{publisher}{ACM}, \bibinfo{address}{New York, NY, USA},
  \bibinfo{pages}{157--168}.
\newblock


\bibitem[\protect\citeauthoryear{L\"{o}ding, Madhusudan, and
  Pe\~{n}a}{L\"{o}ding et~al\mbox{.}}{2017}]%
        {Loding2017}
\bibfield{author}{\bibinfo{person}{Christof L\"{o}ding}, \bibinfo{person}{P.
  Madhusudan}, {and} \bibinfo{person}{Lucas Pe\~{n}a}.}
  \bibinfo{year}{2017}\natexlab{}.
\newblock \showarticletitle{Foundations for Natural Proofs and Quantifier
  Instantiation}.
\newblock \bibinfo{journal}{\emph{Proc. ACM Program. Lang.}}
  \bibinfo{volume}{2}, \bibinfo{number}{POPL}, Article \bibinfo{articleno}{10}
  (\bibinfo{date}{Dec.} \bibinfo{year}{2017}), \bibinfo{numpages}{30}~pages.
\newblock
\showISSN{2475-1421}
\urldef\tempurl%
\url{https://doi.org/10.1145/3158098}
\showDOI{\tempurl}


\bibitem[\protect\citeauthoryear{McKinley}{McKinley}{1991}]%
        {McKinley}
\bibfield{author}{\bibinfo{person}{Kathryn McKinley}.}
  \bibinfo{year}{1991}\natexlab{}.
\newblock \bibinfo{booktitle}{\emph{Dependence Analysis of Arrays Subscriptecl
  by Index Arrays}}.
\newblock \bibinfo{type}{{T}echnical {R}eport} TR91187.
  \bibinfo{institution}{Rice University}.
\newblock


\bibitem[\protect\citeauthoryear{Moon and Hall}{Moon and Hall}{1999}]%
        {Moon:1999}
\bibfield{author}{\bibinfo{person}{Sungdo Moon} {and} \bibinfo{person}{Mary~W.
  Hall}.} \bibinfo{year}{1999}\natexlab{}.
\newblock \showarticletitle{Evaluation of Predicated Array Data-flow Analysis
  for Automatic Parallelization}. In \bibinfo{booktitle}{\emph{Proceedings of
  the Seventh ACM SIGPLAN Symposium on Principles and Practice of Parallel
  Programming}} \emph{(\bibinfo{series}{PPoPP '99})}. \bibinfo{publisher}{ACM},
  \bibinfo{address}{New York, NY, USA}, \bibinfo{pages}{84--95}.
\newblock
\showISBNx{1-58113-100-3}
\urldef\tempurl%
\url{https://doi.org/10.1145/301104.301112}
\showDOI{\tempurl}


\bibitem[\protect\citeauthoryear{Moura and Bj{\o}rner}{Moura and
  Bj{\o}rner}{2007}]%
        {moura2007}
\bibfield{author}{\bibinfo{person}{Leonardo Moura} {and}
  \bibinfo{person}{Nikolaj Bj{\o}rner}.} \bibinfo{year}{2007}\natexlab{}.
\newblock \showarticletitle{Efficient E-Matching for SMT Solvers}. In
  \bibinfo{booktitle}{\emph{Proceedings of the 21st International Conference on
  Automated Deduction: Automated Deduction}}
  \emph{(\bibinfo{series}{CADE-21})}. \bibinfo{pages}{183--198}.
\newblock
\urldef\tempurl%
\url{https://doi.org/10.1007/978-3-540-73595-3_13}
\showDOI{\tempurl}


\bibitem[\protect\citeauthoryear{Oancea and Rauchwerger}{Oancea and
  Rauchwerger}{2012}]%
        {Oancea2012}
\bibfield{author}{\bibinfo{person}{Cosmin~E. Oancea} {and}
  \bibinfo{person}{Lawrence Rauchwerger}.} \bibinfo{year}{2012}\natexlab{}.
\newblock \showarticletitle{Logical inference techniques for loop
  parallelization}. In \bibinfo{booktitle}{\emph{Proceedings of the 33rd ACM
  SIGPLAN conference on Programming Language Design and Implementation}}
  \emph{(\bibinfo{series}{PLDI '12})}. \bibinfo{publisher}{ACM},
  \bibinfo{address}{New York, NY, USA}, \bibinfo{pages}{509--520}.
\newblock


\bibitem[\protect\citeauthoryear{Paek, Hoeflinger, and Padua}{Paek
  et~al\mbox{.}}{2002}]%
        {Paek2002}
\bibfield{author}{\bibinfo{person}{Yunheung Paek}, \bibinfo{person}{Jay
  Hoeflinger}, {and} \bibinfo{person}{David Padua}.}
  \bibinfo{year}{2002}\natexlab{}.
\newblock \showarticletitle{Efficient and Precise Array Access Analysis}.
\newblock \bibinfo{journal}{\emph{ACM Trans. Program. Lang. Syst.}}
  \bibinfo{volume}{24}, \bibinfo{number}{1} (\bibinfo{date}{Jan.}
  \bibinfo{year}{2002}), \bibinfo{pages}{65--109}.
\newblock


\bibitem[\protect\citeauthoryear{Park, Smelyanskiy, Sundaram, and Dubey}{Park
  et~al\mbox{.}}{2014a}]%
        {Park2014}
\bibfield{author}{\bibinfo{person}{Jongsoo Park}, \bibinfo{person}{Mikhail
  Smelyanskiy}, \bibinfo{person}{Narayanan Sundaram}, {and}
  \bibinfo{person}{Pradeep Dubey}.} \bibinfo{year}{2014}\natexlab{a}.
\newblock \showarticletitle{Sparsifying Synchronization for High-Performance
  Shared-Memory Sparse Triangular Solver}. In
  \bibinfo{booktitle}{\emph{Proceedings of the 29th International Conference on
  Supercomputing - Volume 8488}} \emph{(\bibinfo{series}{ISC 2014})}.
  \bibinfo{publisher}{Springer-Verlag New York, Inc.}, \bibinfo{address}{New
  York, NY, USA}, \bibinfo{pages}{124--140}.
\newblock


\bibitem[\protect\citeauthoryear{Park, Smelyanskiy, Vaidyanathan, Heinecke,
  Kalamkar, Liu, Patwary, Lu, and Dubey}{Park et~al\mbox{.}}{2014b}]%
        {Jongsoo14SC}
\bibfield{author}{\bibinfo{person}{Jongsoo Park}, \bibinfo{person}{Mikhail
  Smelyanskiy}, \bibinfo{person}{Karthikeyan Vaidyanathan},
  \bibinfo{person}{Alexander Heinecke}, \bibinfo{person}{Dhiraj~D. Kalamkar},
  \bibinfo{person}{Xing Liu}, \bibinfo{person}{Md. Mosotofa~Ali Patwary},
  \bibinfo{person}{Yutong Lu}, {and} \bibinfo{person}{Pradeep Dubey}.}
  \bibinfo{year}{2014}\natexlab{b}.
\newblock \showarticletitle{Efficient Shared-memory Implementation of
  High-performance Conjugate Gradient Benchmark and Its Application to
  Unstructured Matrices}. In \bibinfo{booktitle}{\emph{Proc. Int. Conf. for
  High Performance Computing, Networking, Storage and Analysis}}
  \emph{(\bibinfo{series}{SC '14})}. \bibinfo{publisher}{IEEE Press},
  \bibinfo{address}{Piscataway, NJ, USA}, \bibinfo{pages}{945--955}.
\newblock


\bibitem[\protect\citeauthoryear{Pothen and Sun}{Pothen and Sun}{1993}]%
        {pothen1993mapping}
\bibfield{author}{\bibinfo{person}{Alex Pothen} {and}
  \bibinfo{person}{Chunguang Sun}.} \bibinfo{year}{1993}\natexlab{}.
\newblock \showarticletitle{A mapping algorithm for parallel sparse Cholesky
  factorization}.
\newblock \bibinfo{journal}{\emph{SIAM Journal on Scientific Computing}}
  \bibinfo{volume}{14}, \bibinfo{number}{5} (\bibinfo{year}{1993}),
  \bibinfo{pages}{1253--1257}.
\newblock


\bibitem[\protect\citeauthoryear{Pothen and Toledo}{Pothen and Toledo}{2004}]%
        {pothen2004elimination}
\bibfield{author}{\bibinfo{person}{Alex Pothen} {and} \bibinfo{person}{Sivan
  Toledo}.} \bibinfo{year}{2004}\natexlab{}.
\newblock \bibinfo{title}{Elimination Structures in Scientific Computing.}
\newblock   (\bibinfo{year}{2004}).
\newblock


\bibitem[\protect\citeauthoryear{Pozo, Remington, and Lumsdaine}{Pozo
  et~al\mbox{.}}{1996}]%
        {pozo1996sparselib++}
\bibfield{author}{\bibinfo{person}{Roldan Pozo}, \bibinfo{person}{Karin
  Remington}, {and} \bibinfo{person}{Andrew Lumsdaine}.}
  \bibinfo{year}{1996}\natexlab{}.
\newblock \showarticletitle{SparseLib++ v. 1.5 Sparse Matrix Class Library
  reference guide}.
\newblock \bibinfo{journal}{\emph{NIST Interagency/Internal Report
  (NISTIR)-5861}} (\bibinfo{year}{1996}).
\newblock


\bibitem[\protect\citeauthoryear{Pugh and Wonnacott}{Pugh and
  Wonnacott}{1994}]%
        {Nonlinear94}
\bibfield{author}{\bibinfo{person}{Bill Pugh} {and} \bibinfo{person}{David
  Wonnacott}.} \bibinfo{year}{1994}\natexlab{}.
\newblock \bibinfo{booktitle}{\emph{Nonlinear Array Dependence Analysis}}.
\newblock \bibinfo{type}{{T}echnical {R}eport} CS-TR-3372.
  \bibinfo{institution}{Dept. of Computer Science, Univ. of Maryland}.
\newblock


\bibitem[\protect\citeauthoryear{Pugh and Rosser}{Pugh and Rosser}{1997}]%
        {Pugh97}
\bibfield{author}{\bibinfo{person}{William Pugh} {and} \bibinfo{person}{Evan
  Rosser}.} \bibinfo{year}{1997}\natexlab{}.
\newblock \showarticletitle{Iteration space slicing and its application to
  communication optimization}. In \bibinfo{booktitle}{\emph{Proceedings of the
  11th international conference on Supercomputing}}. \bibinfo{publisher}{ACM
  Press}, \bibinfo{pages}{221--228}.
\newblock


\bibitem[\protect\citeauthoryear{Pugh and Wonnacott}{Pugh and
  Wonnacott}{1995}]%
        {PughWonn95}
\bibfield{author}{\bibinfo{person}{William Pugh} {and} \bibinfo{person}{David
  Wonnacott}.} \bibinfo{year}{1995}\natexlab{}.
\newblock \showarticletitle{Nonlinear Array Dependence Analysis}. In
  \bibinfo{booktitle}{\emph{Third Workshop on Languages, Compilers, and
  Run-Time Systems for Scalable Computers}}. \bibinfo{address}{Troy, New York}.
\newblock


\bibitem[\protect\citeauthoryear{Pugh and Wonnacott}{Pugh and
  Wonnacott}{1998}]%
        {pugh98constraintbased}
\bibfield{author}{\bibinfo{person}{William Pugh} {and} \bibinfo{person}{David
  Wonnacott}.} \bibinfo{year}{1998}\natexlab{}.
\newblock \showarticletitle{Constraint-Based Array Dependence Analysis}.
\newblock \bibinfo{journal}{\emph{ACM Transactions on Programming Languages and
  Systems}} \bibinfo{volume}{20}, \bibinfo{number}{3} (\bibinfo{date}{1~May}
  \bibinfo{year}{1998}), \bibinfo{pages}{635--678}.
\newblock


\bibitem[\protect\citeauthoryear{Rauchwerger, Amato, and Padua}{Rauchwerger
  et~al\mbox{.}}{1995a}]%
        {Rauchwerger95}
\bibfield{author}{\bibinfo{person}{L. Rauchwerger}, \bibinfo{person}{N.~M.
  Amato}, {and} \bibinfo{person}{D.~A. Padua}.}
  \bibinfo{year}{1995}\natexlab{a}.
\newblock \showarticletitle{Run-Time Methods for Parallelizing Partially
  Parallel Loops}. In \bibinfo{booktitle}{\emph{Proceedings of the {ACM}
  International Conference on Supercomputing (ICS)}}. \bibinfo{publisher}{ACM},
  \bibinfo{address}{New York, NY, USA}, \bibinfo{pages}{137--146}.
\newblock


\bibitem[\protect\citeauthoryear{Rauchwerger, Amato, and Padua}{Rauchwerger
  et~al\mbox{.}}{1995b}]%
        {rauchwerger95scalable}
\bibfield{author}{\bibinfo{person}{Lawrence Rauchwerger},
  \bibinfo{person}{Nancy~M. Amato}, {and} \bibinfo{person}{David~A. Padua}.}
  \bibinfo{year}{1995}\natexlab{b}.
\newblock \showarticletitle{A Scalable Method for Run-Time Loop
  Parallelization}.
\newblock \bibinfo{journal}{\emph{International Journal of Parallel
  Programming}} \bibinfo{volume}{23}, \bibinfo{number}{6}
  (\bibinfo{year}{1995}), \bibinfo{pages}{537--576}.
\newblock


\bibitem[\protect\citeauthoryear{Rauchwerger and Padua}{Rauchwerger and
  Padua}{1999}]%
        {Rauchwerger:1999}
\bibfield{author}{\bibinfo{person}{Lawrence Rauchwerger} {and}
  \bibinfo{person}{David~A. Padua}.} \bibinfo{year}{1999}\natexlab{}.
\newblock \showarticletitle{The LRPD Test: Speculative Run-Time Parallelization
  of Loops with Privatization and Reduction Parallelization}.
\newblock \bibinfo{journal}{\emph{IEEE Trans. Parallel Distrib. Syst.}}
  \bibinfo{volume}{10}, \bibinfo{number}{2} (\bibinfo{date}{Feb.}
  \bibinfo{year}{1999}), \bibinfo{pages}{160--180}.
\newblock
\showISSN{1045-9219}
\urldef\tempurl%
\url{https://doi.org/10.1109/71.752782}
\showDOI{\tempurl}


\bibitem[\protect\citeauthoryear{Ravishankar, Dathathri, Elango, Pouchet,
  Ramanujam, Rountev, and Sadayappan}{Ravishankar et~al\mbox{.}}{2015}]%
        {Ravishankar2015}
\bibfield{author}{\bibinfo{person}{Mahesh Ravishankar}, \bibinfo{person}{Roshan
  Dathathri}, \bibinfo{person}{Venmugil Elango},
  \bibinfo{person}{Louis-No\"{e}l Pouchet}, \bibinfo{person}{J. Ramanujam},
  \bibinfo{person}{Atanas Rountev}, {and} \bibinfo{person}{P. Sadayappan}.}
  \bibinfo{year}{2015}\natexlab{}.
\newblock \showarticletitle{Distributed Memory Code Generation for Mixed
  Irregular/Regular Computations}. In \bibinfo{booktitle}{\emph{Proceedings of
  the 20th ACM SIGPLAN Symposium on Principles and Practice of Parallel
  Programming}} \emph{(\bibinfo{series}{PPoPP 2015})}.
  \bibinfo{publisher}{ACM}, \bibinfo{address}{New York, NY, USA},
  \bibinfo{pages}{65--75}.
\newblock
\showISBNx{978-1-4503-3205-7}
\urldef\tempurl%
\url{https://doi.org/10.1145/2688500.2688515}
\showDOI{\tempurl}


\bibitem[\protect\citeauthoryear{Rennich, Stosic, and Davis}{Rennich
  et~al\mbox{.}}{2016}]%
        {rennich2016accelerating}
\bibfield{author}{\bibinfo{person}{Steven~C Rennich}, \bibinfo{person}{Darko
  Stosic}, {and} \bibinfo{person}{Timothy~A Davis}.}
  \bibinfo{year}{2016}\natexlab{}.
\newblock \showarticletitle{Accelerating sparse Cholesky factorization on
  GPUs}.
\newblock \bibinfo{journal}{\emph{Parallel Comput.}}  \bibinfo{volume}{59}
  (\bibinfo{year}{2016}), \bibinfo{pages}{140--150}.
\newblock


\bibitem[\protect\citeauthoryear{Reynolds, Deters, Kuncak, Barrett, and
  Tinelli}{Reynolds et~al\mbox{.}}{2015}]%
        {reynolds2015counterexample}
\bibfield{author}{\bibinfo{person}{Andrew Reynolds}, \bibinfo{person}{Morgan
  Deters}, \bibinfo{person}{Viktor Kuncak}, \bibinfo{person}{Clark~W. Barrett},
  {and} \bibinfo{person}{Cesare Tinelli}.} \bibinfo{year}{2015}\natexlab{}.
\newblock \showarticletitle{On Counterexample Guided Quantifier Instantiation
  for Synthesis in {CVC4}}.
\newblock \bibinfo{journal}{\emph{CoRR}}  \bibinfo{volume}{abs/1502.04464}
  (\bibinfo{year}{2015}).
\newblock
\urldef\tempurl%
\url{http://arxiv.org/abs/1502.04464}
\showURL{%
\tempurl}


\bibitem[\protect\citeauthoryear{Reynolds, Tinelli, and de~Moura}{Reynolds
  et~al\mbox{.}}{2014}]%
        {reynolds2014finding}
\bibfield{author}{\bibinfo{person}{Andrew Reynolds}, \bibinfo{person}{Cesare
  Tinelli}, {and} \bibinfo{person}{Leonardo de Moura}.}
  \bibinfo{year}{2014}\natexlab{}.
\newblock \showarticletitle{Finding Conflicting Instances of Quantified
  Formulas in SMT}. In \bibinfo{booktitle}{\emph{Proceedings of the 14th
  Conference on Formal Methods in Computer-Aided Design}}
  \emph{(\bibinfo{series}{FMCAD '14})}. Article \bibinfo{articleno}{31},
  \bibinfo{numpages}{8}~pages.
\newblock


\bibitem[\protect\citeauthoryear{Rothberg and Gupta}{Rothberg and
  Gupta}{1992}]%
        {Rothberg92}
\bibfield{author}{\bibinfo{person}{Edward Rothberg} {and}
  \bibinfo{person}{Anoop Gupta}.} \bibinfo{year}{1992}\natexlab{}.
\newblock \showarticletitle{Parallel {ICCG} on a hierarchical memory
  multiprocessor - Addressing the triangular solve bottleneck}.
\newblock \bibinfo{journal}{\emph{Parallel Comput.}} \bibinfo{volume}{18},
  \bibinfo{number}{7} (\bibinfo{year}{1992}), \bibinfo{pages}{719 -- 741}.
\newblock
\urldef\tempurl%
\url{https://doi.org/10.1016/0167-8191(92)90041-5}
\showDOI{\tempurl}


\bibitem[\protect\citeauthoryear{Rus, Hoeflinger, and Rauchwerger}{Rus
  et~al\mbox{.}}{2003}]%
        {Rus2003}
\bibfield{author}{\bibinfo{person}{Silvius Rus}, \bibinfo{person}{Jay
  Hoeflinger}, {and} \bibinfo{person}{Lawrence Rauchwerger}.}
  \bibinfo{year}{2003}\natexlab{}.
\newblock \showarticletitle{Hybrid analysis: static \& dynamic memory reference
  analysis}.
\newblock \bibinfo{journal}{\emph{International Journal Parallel Programming}}
  \bibinfo{volume}{31}, \bibinfo{number}{4} (\bibinfo{year}{2003}),
  \bibinfo{pages}{251--283}.
\newblock


\bibitem[\protect\citeauthoryear{Saltz}{Saltz}{1990}]%
        {Saltz90}
\bibfield{author}{\bibinfo{person}{Joel~H. Saltz}.}
  \bibinfo{year}{1990}\natexlab{}.
\newblock \showarticletitle{Aggregation Methods for Solving Sparse Triangular
  Systems on Multiprocessors}.
\newblock \bibinfo{journal}{\emph{SIAM J. Sci. Stat. Comput.}}
  \bibinfo{volume}{11}, \bibinfo{number}{1} (\bibinfo{date}{Jan.}
  \bibinfo{year}{1990}), \bibinfo{pages}{123--144}.
\newblock
\urldef\tempurl%
\url{https://doi.org/10.1137/0911008}
\showDOI{\tempurl}


\bibitem[\protect\citeauthoryear{Saltz, Mirchandaney, and Crowley}{Saltz
  et~al\mbox{.}}{1991}]%
        {Saltz91}
\bibfield{author}{\bibinfo{person}{Joel~H. Saltz}, \bibinfo{person}{Ravi
  Mirchandaney}, {and} \bibinfo{person}{Kay Crowley}.}
  \bibinfo{year}{1991}\natexlab{}.
\newblock \showarticletitle{Run-Time Parallelization and Scheduling of Loops}.
\newblock \bibinfo{journal}{\emph{IEEE Trans. Comput.}} \bibinfo{volume}{40},
  \bibinfo{number}{5} (\bibinfo{year}{1991}), \bibinfo{pages}{603--612}.
\newblock


\bibitem[\protect\citeauthoryear{Schenk and G{\"a}rtner}{Schenk and
  G{\"a}rtner}{2002}]%
        {schenk2002two}
\bibfield{author}{\bibinfo{person}{Olaf Schenk} {and} \bibinfo{person}{Klaus
  G{\"a}rtner}.} \bibinfo{year}{2002}\natexlab{}.
\newblock \showarticletitle{Two-level dynamic scheduling in PARDISO: Improved
  scalability on shared memory multiprocessing systems}.
\newblock \bibinfo{journal}{\emph{Parallel Comput.}} \bibinfo{volume}{28},
  \bibinfo{number}{2} (\bibinfo{year}{2002}), \bibinfo{pages}{187--197}.
\newblock


\bibitem[\protect\citeauthoryear{Shostak}{Shostak}{1979}]%
        {Shostak79}
\bibfield{author}{\bibinfo{person}{Robert~E. Shostak}.}
  \bibinfo{year}{1979}\natexlab{}.
\newblock \showarticletitle{A Practical Decision Procedure for Arithmetic with
  Function Symbols}.
\newblock \bibinfo{journal}{\emph{J. ACM}} \bibinfo{volume}{26},
  \bibinfo{number}{2} (\bibinfo{date}{April} \bibinfo{year}{1979}),
  \bibinfo{pages}{351--360}.
\newblock
\showISSN{0004-5411}
\urldef\tempurl%
\url{https://doi.org/10.1145/322123.322137}
\showDOI{\tempurl}


\bibitem[\protect\citeauthoryear{Streit, Doerfert, Hammacher, Zeller, and
  Hack}{Streit et~al\mbox{.}}{2015}]%
        {Streit2015}
\bibfield{author}{\bibinfo{person}{Kevin Streit}, \bibinfo{person}{Johannes
  Doerfert}, \bibinfo{person}{Clemens Hammacher}, \bibinfo{person}{Andreas
  Zeller}, {and} \bibinfo{person}{Sebastian Hack}.}
  \bibinfo{year}{2015}\natexlab{}.
\newblock \showarticletitle{Generalized Task Parallelism}.
\newblock \bibinfo{journal}{\emph{ACM Trans. Archit. Code Optim.}}
  \bibinfo{volume}{12}, \bibinfo{number}{1}, Article \bibinfo{articleno}{8}
  (\bibinfo{date}{April} \bibinfo{year}{2015}), \bibinfo{numpages}{25}~pages.
\newblock
\showISSN{1544-3566}
\urldef\tempurl%
\url{https://doi.org/10.1145/2723164}
\showDOI{\tempurl}


\bibitem[\protect\citeauthoryear{Strout, Carter, Ferrante, and Kreaseck}{Strout
  et~al\mbox{.}}{2004}]%
        {StroutIJHPCA}
\bibfield{author}{\bibinfo{person}{Michelle~Mills Strout},
  \bibinfo{person}{Larry Carter}, \bibinfo{person}{Jeanne Ferrante}, {and}
  \bibinfo{person}{Barbara Kreaseck}.} \bibinfo{year}{2004}\natexlab{}.
\newblock \showarticletitle{Sparse Tiling for Stationary Iterative Methods}.
\newblock \bibinfo{journal}{\emph{International Journal of High Performance
  Computing Applications}} \bibinfo{volume}{18}, \bibinfo{number}{1}
  (\bibinfo{date}{February} \bibinfo{year}{2004}), \bibinfo{pages}{95--114}.
\newblock


\bibitem[\protect\citeauthoryear{Strout, LaMielle, Carter, Ferrante, Kreaseck,
  and Olschanowsky}{Strout et~al\mbox{.}}{2016}]%
        {Strout16}
\bibfield{author}{\bibinfo{person}{Michelle~Mills Strout},
  \bibinfo{person}{Alan LaMielle}, \bibinfo{person}{Larry Carter},
  \bibinfo{person}{Jeanne Ferrante}, \bibinfo{person}{Barbara Kreaseck}, {and}
  \bibinfo{person}{Catherine Olschanowsky}.} \bibinfo{year}{2016}\natexlab{}.
\newblock \showarticletitle{An Approach for Code Generation in the Sparse
  Polyhedral Framework}.
\newblock \bibinfo{journal}{\emph{Parallel Comput.}} \bibinfo{volume}{53},
  \bibinfo{number}{C} (\bibinfo{date}{April} \bibinfo{year}{2016}),
  \bibinfo{pages}{32--57}.
\newblock


\bibitem[\protect\citeauthoryear{Venkat, Soltan~Mohammadi, Park, Rong, Barik,
  Strout, and Hall}{Venkat et~al\mbox{.}}{2016}]%
        {Venkat:2016}
\bibfield{author}{\bibinfo{person}{Anand Venkat}, \bibinfo{person}{Mahdi
  Soltan~Mohammadi}, \bibinfo{person}{Jongsoo Park}, \bibinfo{person}{Hongbo
  Rong}, \bibinfo{person}{Rajkishore Barik}, \bibinfo{person}{Michelle~Mills
  Strout}, {and} \bibinfo{person}{Mary Hall}.} \bibinfo{year}{2016}\natexlab{}.
\newblock \showarticletitle{Automating Wavefront Parallelization for Sparse
  Matrix Computations}. In \bibinfo{booktitle}{\emph{Proceedings of the
  International Conference for High Performance Computing, Networking, Storage
  and Analysis}} \emph{(\bibinfo{series}{SC '16})}. \bibinfo{publisher}{IEEE
  Press}, \bibinfo{address}{Piscataway, NJ, USA}, Article
  \bibinfo{articleno}{41}, \bibinfo{numpages}{12}~pages.
\newblock
\showISBNx{978-1-4673-8815-3}
\urldef\tempurl%
\url{http://dl.acm.org/citation.cfm?id=3014904.3014959}
\showURL{%
\tempurl}


\bibitem[\protect\citeauthoryear{Verdoolaege}{Verdoolaege}{2010}]%
        {isl10}
\bibfield{author}{\bibinfo{person}{Sven Verdoolaege}.}
  \bibinfo{year}{2010}\natexlab{}.
\newblock \showarticletitle{isl: An integer set library for the polyhedral
  model}. In \bibinfo{booktitle}{\emph{Proceedings of the 3rd International
  Congress on Mathematical Software}} \emph{(\bibinfo{series}{ICMS '10})}.
  \bibinfo{pages}{299--302}.
\newblock


\bibitem[\protect\citeauthoryear{Verdoolaege}{Verdoolaege}{2018}]%
        {isl2018}
\bibfield{author}{\bibinfo{person}{Sven Verdoolaege}.}
  \bibinfo{year}{2018}\natexlab{}.
\newblock \bibinfo{booktitle}{\emph{Integer Set Library: Manual}}.
\newblock
\urldef\tempurl%
\url{http://isl.gforge.inria.fr//manual.pdf}
\showURL{%
\tempurl}


\bibitem[\protect\citeauthoryear{Vuduc, Kamil, Hsu, Nishtala, Demmel, and
  Yelick}{Vuduc et~al\mbox{.}}{2002}]%
        {vuduc2002automatic}
\bibfield{author}{\bibinfo{person}{Richard Vuduc}, \bibinfo{person}{Shoaib
  Kamil}, \bibinfo{person}{Jen Hsu}, \bibinfo{person}{Rajesh Nishtala},
  \bibinfo{person}{James~W Demmel}, {and} \bibinfo{person}{Katherine~A
  Yelick}.} \bibinfo{year}{2002}\natexlab{}.
\newblock \showarticletitle{Automatic performance tuning and analysis of sparse
  triangular solve}. ICS.
\newblock


\bibitem[\protect\citeauthoryear{Wang, Zhang, Shen, Zhang, Lu, Wu, and
  Wang}{Wang et~al\mbox{.}}{2014}]%
        {wang2014intel}
\bibfield{author}{\bibinfo{person}{Endong Wang}, \bibinfo{person}{Qing Zhang},
  \bibinfo{person}{Bo Shen}, \bibinfo{person}{Guangyong Zhang},
  \bibinfo{person}{Xiaowei Lu}, \bibinfo{person}{Qing Wu}, {and}
  \bibinfo{person}{Yajuan Wang}.} \bibinfo{year}{2014}\natexlab{}.
\newblock \showarticletitle{Intel math kernel library}.
\newblock In \bibinfo{booktitle}{\emph{High-Performance Computing on the
  Intel{\textregistered} Xeon Phi}}. \bibinfo{publisher}{Springer},
  \bibinfo{pages}{167--188}.
\newblock


\bibitem[\protect\citeauthoryear{Wolfe}{Wolfe}{1989}]%
        {Wolfe:1989}
\bibfield{author}{\bibinfo{person}{M. Wolfe}.} \bibinfo{year}{1989}\natexlab{}.
\newblock \showarticletitle{More Iteration Space Tiling}. In
  \bibinfo{booktitle}{\emph{Proceedings of the 1989 ACM/IEEE Conference on
  Supercomputing}} \emph{(\bibinfo{series}{Supercomputing '89})}.
  \bibinfo{publisher}{ACM}, \bibinfo{address}{New York, NY, USA},
  \bibinfo{pages}{655--664}.
\newblock
\showISBNx{0-89791-341-8}
\urldef\tempurl%
\url{https://doi.org/10.1145/76263.76337}
\showDOI{\tempurl}


\bibitem[\protect\citeauthoryear{Zheng, Rogers, Luo, Dwyer, and Siegel}{Zheng
  et~al\mbox{.}}{2015a}]%
        {Zheng:2015}
\bibfield{author}{\bibinfo{person}{M. Zheng}, \bibinfo{person}{M.~S. Rogers},
  \bibinfo{person}{Z. Luo}, \bibinfo{person}{M.~B. Dwyer}, {and}
  \bibinfo{person}{S.~F. Siegel}.} \bibinfo{year}{2015}\natexlab{a}.
\newblock \showarticletitle{CIVL: Formal Verification of Parallel Programs}. In
  \bibinfo{booktitle}{\emph{2015 30th IEEE/ACM International Conference on
  Automated Software Engineering (ASE)}}. \bibinfo{pages}{830--835}.
\newblock
\urldef\tempurl%
\url{https://doi.org/10.1109/ASE.2015.99}
\showDOI{\tempurl}


\bibitem[\protect\citeauthoryear{Zheng, Wang, Jin, Wu, Chen, and Jiang}{Zheng
  et~al\mbox{.}}{2015b}]%
        {zheng2015gpu}
\bibfield{author}{\bibinfo{person}{Ran Zheng}, \bibinfo{person}{Wei Wang},
  \bibinfo{person}{Hai Jin}, \bibinfo{person}{Song Wu}, \bibinfo{person}{Yong
  Chen}, {and} \bibinfo{person}{Han Jiang}.} \bibinfo{year}{2015}\natexlab{b}.
\newblock \showarticletitle{GPU-based multifrontal optimizing method in sparse
  Cholesky factorization}. In \bibinfo{booktitle}{\emph{Application-specific
  Systems, Architectures and Processors (ASAP), 2015 IEEE 26th International
  Conference on}}. IEEE, \bibinfo{pages}{90--97}.
\newblock


\bibitem[\protect\citeauthoryear{Zhuang, Eichenberger, Luo, O'Brien, and
  O'Brien}{Zhuang et~al\mbox{.}}{2009}]%
        {Zhuang09}
\bibfield{author}{\bibinfo{person}{Xiaotong Zhuang}, \bibinfo{person}{A.E.
  Eichenberger}, \bibinfo{person}{Yangchun Luo}, \bibinfo{person}{K. O'Brien},
  {and} \bibinfo{person}{K. O'Brien}.} \bibinfo{year}{2009}\natexlab{}.
\newblock \showarticletitle{Exploiting Parallelism with Dependence-Aware
  Scheduling}. In \bibinfo{booktitle}{\emph{International Conference on
  Parallel Architectures and Compilation Techniques (PACT)}}.
  \bibinfo{publisher}{IEEE Computer Society}, \bibinfo{address}{Los Alamitos,
  CA, USA}, \bibinfo{pages}{193--202}.
\newblock


\end{thebibliography}



\end{document}